%% file: main.tex
\documentclass{article}
\usepackage[]{graphicx}
\usepackage[]{xcolor}

\makeatletter
\def\maxwidth{ %
  \ifdim\Gin@nat@width>\linewidth
    \linewidth
  \else
    \Gin@nat@width
  \fi
}
\makeatother

\definecolor{fgcolor}{rgb}{0.345, 0.345, 0.345}

\usepackage{framed}
\makeatletter
 {\par\unskip\endMakeFramed%
 \at@end@of@kframe}
\makeatother

\definecolor{shadecolor}{rgb}{.97, .97, .97}
\definecolor{messagecolor}{rgb}{0, 0, 0}
\definecolor{warningcolor}{rgb}{1, 0, 1}
\definecolor{errorcolor}{rgb}{1, 0, 0}

\usepackage{alltt}
\usepackage{pdfpages}
\usepackage{bbm}
\usepackage{mathtools} 
\usepackage{geometry}
\usepackage{url}
\usepackage[unicode=true,pdfusetitle,
 bookmarks=true,bookmarksnumbered=true,
 bookmarksopen=true,bookmarksopenlevel=2,
 breaklinks=false,pdfborder={0 0 1},backref=false, colorlinks=false, hidelinks]
 {hyperref}
\hypersetup{
 pdfstartview={XYZ null null 1}}
\usepackage{cleveref}
\crefformat{equation}{#2equation~#1#3}
\Crefformat{equation}{#2Equation~#1#3}
\usepackage{amsmath,amssymb}
\usepackage{breakurl}
\usepackage{tocloft}
\usepackage{amsmath}
\usepackage{booktabs,caption,fixltx2e}
\usepackage[flushleft]{threeparttable}
\usepackage{bm}
\usepackage{subcaption}
\usepackage[english]{babel}
\usepackage{csquotes}
\usepackage[style=numeric-comp, sorting=none, backend=biber]{biblatex}

\newcommand\independent{\protect\mathpalette{\protect\independenT}{\perp}}
\def\independenT#1#2{\mathrel{\rlap{$#1#2$}\mkern2mu{#1#2}}}
\usepackage{titlesec} 
\bibliography{comp_methods}

\usepackage{float}
\usepackage{xcolor}
\usepackage{setspace}
\usepackage{comment}
\usepackage{authblk}

\usepackage{subfiles} 

\author[1,*]{Alejandra Benitez}
\affil{Genentech Inc.,
South San Francisco, CA, USA}
\author{Maya L. Petersen}
\author{Mark J. van der Laan}
\affil{University of California Berkeley School of Public Health, Biostatistics, Berkeley, CA, USA}
\author{Nicole Santos}
\author{Elizabeth Butrick}
\author{Dilys Walker}
\author{Rakesh Ghosh}
\affil{University of California San Francisco, Institute for Global Health Sciences, San Francisco, CA, USA}
\author{Phelgona Otieno}
\affil{Kenya Medical Research Institute, Center for Clinical Research, Nairobi, Kenya}
\author{Peter Waiswa}
\affil{Makerere University College of Health Sciences, Centre of Excellence for Maternal, Newborn and Child Health, Kampala, Uganda}
\author[2]{Laura B. Balzer}
\date{\today}

\affil[*]{Corresponding author: Alejandra Benitez\\ Genentech Inc.,
South San Francisco, CA, USA\\alelizbenitez@gmail.com}

\title{Defining and Estimating Effects in Cluster Randomized Trials:\\ A Methods Comparison}
\IfFileExists{upquote.sty}{\usepackage{upquote}}{}
\begin{document}

\maketitle

\begin{abstract}
Across research disciplines, cluster randomized trials (CRTs) are commonly implemented to evaluate interventions delivered to groups of participants, such as communities and clinics. 
Despite advances in the design and analysis of CRTs, several challenges remain.
First, there are many possible ways to specify the causal effect of interest (e.g., at the individual-level or at the cluster-level). Second, the theoretical and practical performance of common methods for CRT analysis remain poorly understood. 
Here, we present a general framework to formally define an array of causal effects in terms of summary measures of counterfactual outcomes.
Next, we provide a comprehensive overview of CRT estimators, including the t-test, generalized estimating equations (GEE), augmented-GEE, and targeted maximum likelihood estimation (TMLE). Using finite sample simulations, we illustrate the practical performance of these estimators for different causal effects and when, as commonly occurs, there are limited numbers of clusters of different sizes. Finally, our application to data from the Preterm Birth Initiative (PTBi) study demonstrates the real-world impact of varying cluster sizes and targeting effects at the cluster-level or at the individual-level.
{Specifically, the relative effect of the PTBI intervention was 0.81 at the cluster-level, corresponding to a 19\% reduction in outcome incidence, and was 0.66 at the individual-level, corresponding to a 34\% reduction in outcome risk.}
Given its flexibility to estimate a variety of  user-specified effects and ability to adaptively adjust for covariates for precision gains while maintaining Type-I error control, we conclude TMLE is a promising tool for CRT analysis.

\end{abstract}


\noindent Short title: Comparative Methods for the Analysis of Cluster Randomized Trials

\newpage 
\section{Introduction}

Cluster randomized trials (CRTs) provide an opportunity to assess the population-level effects of interventions that are randomized to groups of individuals, such as communities, clinics, or schools. These groups are commonly called ``clusters''.
The choice to randomize clusters, instead of individuals, is often driven by the type of intervention as well as practical considerations \cite{moultonhayes}. For example, interventions to improve medical practices are often randomized at the hospital or clinic-level to reduce logistical burden and to minimize potential contamination between arms if individual patients were instead randomized.
The design and conduct of CRTs has improved considerably \cite{turner17a, turner17b, crepsi, Murray2020}, and results from CRTs have been widely published in public health, education, policy, and economics literature \cite{icc}.  However, a recent review found that only 50\% of CRTs were analyzed with appropriate methods \cite{murray}.

Due to the hierarchical nature of the data and the correlation of participant outcomes within clusters, the analysis of CRTs is fundamentally more complicated than for individually randomized trials  \cite{moultonhayes, murray}. To start, there are many ways to define the causal effect of interest in CRTs. We may, for example, be interested in the effect for the sample enrolled in the CRT or for a wider target population. Furthermore, we may be interested in the effect at the individual-level or  cluster-level.
{\color{black}As detailed below, the individual- and cluster-level effects can diverge markedly  under ``informative cluster size'' \cite{Seaman2014, nevail_ics2014}, occurring 
{when cluster size modifies the intervention effect.}
Finally, various summary measures of arm-specific outcomes (e.g., weighted means) and their contrasts (e.g, difference or ratio) may be of interest.}
Altogether, the  causal effect should be determined by the study's goal and primary research question \cite{Petersen2014roadmap, rose_tl, BalzerMLcomm2021, Kahan2021}. This is in line with the International Council for Harmonization (ICH)-E9(R1) guidance for trial protocols to explicitly state the ``estimand'', including the target population, comparison conditions, endpoint, and summary measure \cite{ICHE9}. Adhering to this guidance ensures the statistical estimator follows from the target estimand, which follows from the trial's objective. Neglecting this guidance risks letting the statistical approach determine the effect estimated and thus determine which research question is answered.

 For the analysis of CRTs, statistical estimation and inference must respect that the cluster is the independent unit. Ignoring the dependence of observations within a cluster can lead to power calculations based on the wrong sample size and underestimates of the standard error. Analyses ignoring clustering may also inappropriately attribute the impact of a cluster-level covariate to the intervention and, together with variance underestimation, result in inflated Type-I error rates. Many analytic approaches are available to account for the dependence within a cluster; examples include conducting a cluster-level analysis or applying a correction factor \cite{moultonhayes, murray, turner17a, turner17b, crepsi}.  Once we have committed to an analytic approach that aligns with our research question and addresses clustering,  the adjustment of baseline covariates 
is often considered to
{\color{black}improve precision and, thereby, statistical power. In an individually randomized trial setting, Benkeser \emph{et al.} recently demonstrated an 18\% savings in sample size that could be achieved by including baseline covariates in the analysis \cite{Benkeser2021}. Likewise, using real data from a CRT, Balzer et al. recently showed the adjusted analysis was 5-times more efficient than an unadjusted analysis for the same parameter \cite{Balzer2021twostage}. 
While our focus is on the potential of covariate adjustment to improve precision, we note that in other settings covariate adjustment may be essential to reducing bias due to missingness, selection, or restricted randomization \cite{diaz_missing, fiero, Balzer2020Supp, Balzer2021twostage, Leyrat2013, Leyrat2014, Li2022, NugentTB2022, Li2016, Li2017}.

 Fortunately, many methods are available to incorporate covariates for improved efficiency when estimating intervention effects in CRTs.}  Examples include well-established methods, such as generalized estimating equations (GEE) and  covariate adjusted residuals estimator (CARE), as well as more recent developments, such as  targeted maximum likelihood estimation (TMLE) and Augmented-GEE (Aug-GEE) \cite{liang-zeger, gail-care, moultonhayes, rose_tl, hierarchical, adaptive_prespec,  aug-gee}.
While these methods differ in their exact implementation, each aims to improve the CRT's statistical power by controlling for individual- or cluster-level covariates when fitting the ``outcome regression": the conditional expectation of the outcome given the randomized intervention and the adjustment variables.
{\color{black}As detailed below, these algorithms naturally estimate distinct causal effects, again highlighting the importance of pre-specifying the target estimand and then choosing the optimal estimator of that effect.}
Previous literature has used simulation studies to compare the attained power and Type-I error rate of various approaches to CRT analysis (e.g., \cite{crt_comparative_bellamy, Ding2021}). 
However, to the best of our knowledge, these comparisons have largely excluded  the more recent approaches of  Aug-GEE and TMLE.

This paper aims to provide a general framework for defining causal effects in CRTs and to assess the comparative performance of analytic methods for estimation of those effects.
Building on prior work in CRT analysis (e.g., \cite{Leyrat2014, Li2016, Li2017, imai2009,Cook2016, moultonhayes, Murray2020, Ding2021}), our key contributions are as follows. First, we demonstrate the utility of a non-parametric structural causal model, accounting for clustering, to derive counterfactuals and define a variety of causal effects of potential interest.
Using this causal model framework, we examine the impact of varying cluster sizes when defining the target causal effect and discuss identification of each causal effect as a function of the observed data distribution (i.e., a statistical estimand). Next, {\color{black}we review recently developed CRT estimators along with well-established ones}, emphasizing their natural target of inference. Specifically, we describe each algorithm's ability to estimate marginal effects on the additive or relative scale and at the individual- or cluster-level, while also adjusting for covariates to improve precision. 
To the best of our knowledge, this is the first paper to describe several implementations of TMLE for the analysis of CRTs; these TMLEs  allow for estimation of individual-level effects with cluster-level data and, conversely, for estimation of cluster-level effects with individual-level data. Also to the best of our knowledge, this is the first paper using finite sample simulations to provide a head-to-head comparison of these TMLEs, overall and under informative cluster size. Importantly, this work is motivated by a real data application with 20 clusters of widely variable size.

{\color{black}For an alternative and complementary presentation using  potential outcomes and a ``design-based'' perspective, we refer the reader to Su and Ding \cite{Ding2021}, 
{who focus} on effects defined for the sample and on the absolute scale (i.e., difference in mean outcomes) as well as estimation using weighted least squares regression. Here, we define and estimate effects on both the absolute and relative scale and for both the sample population and the larger target population. Additionally, our work is applicable regardless of the outcome-type (i.e., binary, count, or continuous) and to trials with more complex designs (e.g., pair-matched). While prior work has focused on asymptotic theory and provided simulation results with a fairly large number of clusters, here, we focus on the practical application of these estimators and evaluate their performance in simulations with limited numbers of clusters.}

As motivating example, we consider the Preterm Birth Initiative (PTBi) study, a maternal-infant CRT which took place in 20 health facilities across Kenya and Uganda (ClinicalTrials.gov: NCT03112018). The trial assessed whether a  facility-based intervention, designed to improve the uptake of evidence-based practices, was effective in reducing 28-day mortality among preterm infants. {\color{black}An important feature of PTBi was the widely varying cluster sizes; specifically, the median number of mother-infant dyads {\color{black} within a facility} was 236 and ranged from 29 to 447.} Following PTBi, we focus on a setting where individual participants are grouped into clusters  (e.g., patients in a health facility). However, our discussion and results are equally applicable to other hierarchical data structures that may be observed in a CRT. Examples include households within a community, classrooms within a school, and hospitals within a district.

The remainder of the paper is organized as follows. In Section 2, 
{\color{black}we use a non-parametric structural equation model to provide a broad approach to formally defining causal effects in CRTs.}
We discuss how the primary endpoint may be defined at the individual-level or at the cluster-level, often as an aggregate of individual-level outcomes. 
We highlight how these endpoints can differ in terms of interpretation and magnitude, especially under the setting of informative cluster size. We additionally address the distinction between effects defined for the study sample versus a target population. 
In Section 3, we discuss several CRT estimation methods. For each method, we describe its target of inference and its use of baseline covariates to improve statistical precision.
In Section 4, we provide two simulation studies to  evaluate the finite sample performance of common CRT estimators and to demonstrate the distinction between different causal effects.  In Section 5, we apply these methods to estimate intervention effects for the PTBi study {\color{black} and highlight the real-world impact of varying cluster sizes}. We conclude in Section 6 with a brief discussion. \par

\section{Defining Causal Effects in CRTs}

We begin by formalizing the notation that  will be used throughout. 
{\color{black}We index the cluster, such as a community or hospital, by $j={1,...,J}$.} 
{\color{black} The study clusters could be randomly sampled from a larger target population of clusters or, due to practical constraints, selected for convenience from a set of candidate clusters. In either case, there is a real or theoretical target population of clusters to which we \emph{may} want to make inferences. Below, we explicitly discuss the definition and estimation of the population versus sample effects. In all settings, the selection of clusters should be reflected in the CONSORT diagram.} 
Each cluster  is comprised of a finite set of individuals (i.e., participants), which are indexed $i={1,...,N_j}$. 
The cluster size, denoted $N_j$, could be constant across clusters, but more often varies by cluster.
{\color{black} The $N_j$ study participants could be randomly sampled within a cluster or be a census of all persons in a cluster. Discussion of more complex settings with systematic sampling or other forces of selection bias is important but beyond the scope of this paper \cite{ Leyrat2013, Leyrat2014,  Balzer2021twostage, Li2022,NugentTB2022}. Here, the study participants are representative  of or comprise the cluster.  Thus, for a given cluster, $N_j$ is fixed and finite.}

In each cluster, $\mathbf{W}_j=(W_{1j},\ldots, W_{N_jj})$ denotes the set of baseline covariates for the  study participants.  Additional baseline covariates related to the cluster are denoted $E_j$; these could be summaries of individual-level characteristics or  may have no individual-level counterpart \cite{hierarchical}. 
{\color{black}For ease of notation, we include the cluster size $N_j$, a random variable, in $E_j$.}
{These cluster-level and individual-level covariates are assumed not to be impacted by the intervention, which is randomized to clusters in a CRT.}
In the case of a pair-matched trial, which may increase study power \cite{moultonhayes, imai2009, balzer_matching}, randomization occurs within pairs of clusters matched on characteristics expected to be predictive of the primary endpoint. 
Throughout, we will use $A_j$ as an indicator that cluster $j$ was randomized to the intervention.  The outcome of interest is $\mathbf{Y}_j=(Y_{1j},\ldots, Y_{N_jj})$, which for simplicity we assume is measured for all participants in cluster $j$.
Extensions to handling missing data {time-dependent covariates, censoring}, and selection bias are important, but beyond the scope of this paper \cite{diaz_missing, fiero, Balzer2021twostage, Leyrat2013, Leyrat2014, Li2022, NugentTB2022}.

\subsection{Hierarchical Structural Causal Model}

We now use Pearl's non-parametric structural equation  model \cite{pearl} to represent the hierarchical data generating process of a CRT. 
As detailed in Balzer et al. \cite{hierarchical}, the following model encodes independence between clusters, but makes no restrictions on the dependence of participants within a cluster.
\begin{align}
E&=f_E(U_E)\nonumber\\
\mathbf{W}&=f_\mathbf{W}(E,U_\mathbf{W})\nonumber\\
A&=f_A(U_A)\nonumber\\
\mathbf{Y}&=f_\mathbf{Y}(E,\mathbf{W},A,U_\mathbf{Y})
\label{Eq:GeneralSCM}
\end{align}
Here, $(f_E, f_\mathbf{W} , f_A, f_\mathbf{Y}$) denote the set of functions that determine the values of the {measured} random variables: the cluster-level covariates $E$, the individual-level covariate matrix $\mathbf{W}$, the {cluster-level} intervention assignment $A$, and the vector of individual-level outcomes $\mathbf{Y}$. 
{\color{black}These functions also account for unmeasured factors $(U_E,U_\mathbf{W}, U_A, U_\mathbf{Y})$. By design in a CRT, the unmeasured factors determining the intervention assignment $U_A$ are independent of others.
These functions are also non-parametric with the exception of $f_A$, the function to generate the cluster-level intervention. For example, in a two-armed trial with equal allocation probability, we have $f_A=\mathbb{I}(U_A < 0.5)$ with $U_A \sim Unif(0,1)$.
In contrast, the function to generate the outcome vector $f_{\mathbf{Y}}$ is left unspecified and can take any form of the ``parent'' variables $(E, \mathbf{W}, A)$ and unmeasured factors $U_\mathbf{Y}$. Importantly, the cluster size $N$, included in $E$, can influence the outcome vector $\mathbf{Y}$ in complex ways. For example, very large and very small clusters may have poorer outcomes. Additionally the intervention effect maybe attenuated or, alternatively, enhanced in the largest and smallest clusters. 
Beyond the cluster-level covariates and intervention $(E,A)$, there are additional unmeasured and measured factors influencing and inducing dependence in the outcomes $\mathbf{Y}$ in a cluster.}
For example, the joint error term $U_\mathbf{Y}$ induces correlation among participants' outcomes within a cluster. Additionally,  within cluster $j$, an individual's outcome $Y_{ij}$ may depend on the covariates of others in the same cluster $\mathbf{W}_j$.
{\color{black}In other settings, the dependence between participants in a cluster might be more restricted; for further details, we refer to reader to Balzer et al. \cite{hierarchical}.}

We now consider the data generating process for the PTBi study{\color{black}, where the cluster corresponds to a health facility and the ``individual'' to a mother-infant dyad.}
For cluster $j$, we measure facility-level baseline characteristics $E_j$, including the average monthly delivery volume, facility preparedness assessment score, staff to delivery ratio, and community-type (i.e., urban versus rural).
The facility is then randomly assigned to intervention $(A_j=1)$ or control $(A_j=0)$.  When a mother  
delivers her infant, the covariates for the mother-infant dyad $W_{ij}$ are collected.  
These include the mother's characteristics, such as age, parity, and 
{receiving a cesarean section (C-section)}, and the infant's characteristics, such as sex, weight, length, and 
arm circumference. 
{Again, we assume these covariates $W_{ij}$ are not impacted by the intervention $A$, even though they are measured after randomization.}
Finally, the infant's vital status is recorded; $Y_{ij}$ is an indicator of infant death within 28 days. 
Over the course of study follow-up, we observe many such deliveries, but for the primary population of interest, we restrict to $N_j$ pre-term births, defined as born before 37 weeks of gestation. This process is repeated for the sample size of $J=20$ facilities.

\subsection{Counterfactuals and  Target Causal Effects}

\label{Section:effects}

We generate  counterfactual outcomes by replacing the structural equation $f_A$ in causal model (Eq.~\ref{Eq:GeneralSCM})  with our desired intervention \cite{pearl}. 
Let $Y_{ij}(a)$ be the counterfactual outcome for individual $i$ in cluster $j$ if, possibly contrary-to-fact, their cluster received treatment-level $A=a$. In the PTBi study, $Y_{ij}(1)$ represents the 28-day vital status for infant $i$ in facility $j$ if that facility had been randomized to the intervention arm ($A_j=1$), while $Y_{ij}(0)$ represents the 28-day vital status for infant $i$ in facility $j$ if that facility had been randomized to the control arm ($A_j=0$).

We can also define counterfactual outcomes at the cluster-level by taking aggregates of the individual-level ones. 
{\color{black}Many such summary measures are possible. In line with common practice \cite{moultonhayes,hierarchical, Cook2016, Imbens2004, imai_2008, imai2009, Ding2021},} we focus on weighted sums of  the $N_j$ participants from cluster $j$: 
 \begin{align}
Y_j^c(a) \equiv \sum_{i=1}^{N_j}\alpha_{ij}Y_{ij}(a)
\label{Eq:ClusterCF}
\end{align}
Often, this weight is selected to be the inverse of the cluster size and, thus, constant across participants in a cluster: $\alpha_{ij}=1/N_j$ for $i=\{1,\ldots, N_j\}$. With this choice,  $Y^c_j(a)$ is the average counterfactual outcome for the $N_j$ participants in cluster $j$. For a binary individual-level outcome, using the inverse cluster size  yields a cluster-level outcome $Y^c(a)$ corresponding to a proportion or probability. 
In PTBi, for example, $Y^c_j(a)$ is the counterfactual cumulative incidence of death by 28-days if, possibly contrary-to-fact, facility $j$ received treatment-level $A_j=a$.
Of course, other weighting schemes can be used to summarize the individual-level counterfactual outcomes to the cluster-level.

By applying a summary measure to the distribution of counterfactual outcomes, we define the causal effect corresponding to our research query.
A wide variety of causal parameters can be expressed as the empirical mean over the sample \cite{neyman1923, rubin1990, imbens_2004, imai_2008,  balzer_sate, Cook2016}:  
\begin{align}
\Phi^{c,J}(a) &=\frac{1}{J} \sum_{j=1}^J \left( \sum_{i=1}^{N_j} \alpha_{ij} Y_{ij}(a) \right) =\frac{1}{J} \sum_{j=1}^J Y^c_j(a) 
\label{Eq:Meta}
\end{align}
where $\alpha_{ij}$ again is the user-specified weight. Throughout, superscript $c$ denotes a summary of cluster-level outcomes, and superscript $J$ denotes a sample parameter. 
As the number of clusters grows ($J\rightarrow \infty)$, the sample parameter converges to the expectation over the target population of clusters: 
\begin{align}
  \Phi^c(a)= \mathbb{E}[Y^c(a)]
  \label{Eq:ClusterCausal}
\end{align}
In words, $\Phi^c(a)$ is the expected cluster-level counterfactual  outcome if all clusters in the population had been assigned to treatment-level $A=a$, whereas  $\Phi^{c,J}(a)$ in Eq.~\ref{Eq:Meta} is the average cluster-level counterfactual outcome for the $J$ clusters in the CRT.
For the PTBi study and weights $\alpha_{ij}=1/N_j$,  $\Phi^c(a)$  represents the expected incidence of 28-day mortality among preterm infants if all health facilities in the target population had been assigned to treatment arm $A=a$, while $\Phi^{c,J}(a)$ represents the average incidence if the $J=20$ health facilities included in the study had been assigned to treatment arm $A=a$. 

By taking contrasts of these treatment-specific causal parameters, we can define causal effects on any scale of interest.  For example, we may be interested in the relative effect for the $J$ clusters included in the study: $\Phi^{c,J}(1) \div \Phi^{c,J}(0)$. Alternatively, we may be interested in their difference at the population-level (a.k.a., the average treatment effect): $\Phi^c(1)-\Phi^c(0)$. For simplicity we refer to contrasts defined using Eq.~\ref{Eq:Meta} or~\ref{Eq:ClusterCausal} as ``cluster-level effects".

Letting $N_T\equiv \sum_j N_j$ be the total number of participants in the CRT, we can also use Eq.~\ref{Eq:Meta} to  define causal effects at the individual-level by setting $\alpha_{ij}=J/N_T$:  
\begin{align}
\Phi^{J}(a) &=\frac{1}{J} \sum_{j=1}^J \sum_{i=1}^{N_j} \left( \frac{J}{N_T} Y_{ij}(a) \right) =\frac{1}{N_T} \sum_{j=1}^J \sum_{i=1}^{N_j} Y_{ij}(a)
\label{Eq:IndvSample}
\end{align}
As sample size grows {\color{black}($J \rightarrow \infty$)}, this  converges to the expectation  over the target population of clusters, each containing a finite number of participants:
\begin{align}
\Phi(a) = \mathbb{E}[Y(a)]
\label{Eq:IndvCausal}
\end{align} 
In words, $\Phi(a)$ is the expected individual-level outcome if all clusters in the target population received treatment-level $A=a$, whereas $\Phi^{J}(a)$ in Eq.~\ref{Eq:IndvSample} is  the average individual-level outcome for the $N_T$ participants in the CRT.
In the PTBi study, $\Phi(a)$ represents the counterfactual risk of mortality for a preterm infant if all health facilities  in the target population received treatment-level $A=a$, while $\Phi^J(a)$ represents the counterfactual proportion of preterm infants who would die if the $J=20$ health facilities included in the study had been assigned to treatment arm $A=a$. 
As before, we can take the difference or ratio of these treatment-specific parameters to define causal effects. 
For simplicity we refer to contrasts defined using Eq.~\ref{Eq:IndvSample} or ~\ref{Eq:IndvCausal} as ``individual-level effects".

In summary, we can define a wide variety of causal parameters by considering alternative summary measures of the individual-level or cluster-level counterfactual outcomes. Further generalizations are available in Appendix A of the Supplementary Materials.
When the cluster size varies (i.e., $N_j \ne n, \forall j$), causal parameters giving equal weight to clusters (e.g., Eq.~\ref{Eq:Meta}) will generally differ from causal parameters giving equal weight to participants (e.g., Eq.~\ref{Eq:IndvSample}).
{\color{black}As a toy example, suppose we have $J=5$ clusters of varying sizes, specifically $N_j=10$ for $j=\{1,\ldots,4\}$ and $N_5=10000$. Further suppose the counterfactual probability of the individual-level outcome depends on cluster size, such that the total number of outcomes ($\sum_i^{N_j}Y_{ij}(a)$) is 2 for clusters $j=\{1,\ldots, 4\}$ and 7500 for cluster $j=5$. Then using the inverse cluster size as weight $(\alpha_{ij}=1/N_j$), the cluster-level sample parameter, given in Eq.~\ref{Eq:Meta}, would be $\Phi^{c,J}(a)=(2/10+2/10+2/10+2/10 + 7500/10000)\div 5=0.31$. In contrast, the individual-level sample parameter, given in Eq.~\ref{Eq:IndvSample}, would be $\Phi^J(a)= (2+2+2+2+7500)\div(10+10+10+10+10000)= 0.748$. While this is an extreme example, it illustrates the potential divergence of the treatment-specific means when cluster size varies.}

Additionally since causal effects are defined through contrasts of these treatment-specific means,  effects defined at the cluster-level or individual-level can also be meaningfully different when cluster size varies. 
The potential divergence in these causal effects is exacerbated under informative cluster size, occurring when the intervention effect is modified by cluster size  \cite{Seaman2014, nevail_ics2014}. This scenario is explored in detail in the second simulation study and real data example, below. Nonetheless, it is worth emphasizing that even if the cluster size is constant (i.e., $N_j=n, \forall j$), the cluster-level and individual-level effects have subtly different interpretations, as discussed previously. {In observational studies, failing to recognize that effects defined at the aggregate-level (i.e., at the cluster-level) may differ from effects defined at the individual-level is known as the ``ecological fallacy'' \cite{RothmanModern}.}

For completeness in the above, we defined both sample-specific and population-level measures. 
For ease of comparison of analytic methods, we focus on the population-level effects (Eqs.~\ref{Eq:ClusterCausal} and~\ref{Eq:IndvCausal}) for the remainder of the paper and refer the reader to \cite{neyman1923, rubin1990, imbens_2004, imai_2008, balzer_sate, Balzer2017} for further discussion on the trade-offs between targeting population versus sample effects. In brief, the sample effect might be more appealing when clusters are selected {for convenience} and can be estimated more precisely than the population effect.

\subsection{Observed Data and Identification of Causal Effects}

For a given cluster, the observed data are the set of measured cluster-level covariates, the matrix of individual-level covariates, the randomized treatment, and the vector of individual-level outcomes: $$O=(E,\mathbf{W},A,\mathbf{Y}).$$
We assume the observed data are generated by sampling $J$ times from a data generating process compatible with
{\color{black}the above causal model (Eq.~\ref{Eq:GeneralSCM}).}
This provides a link between the causal model and the statistical model, which is the set of possible distributions of the observed data \cite{rose_tl}. 
The causal model  encodes that the cluster-level treatment is randomized{\color{black}, but does not otherwise place restrictions on the joint distribution of observed data, denoted $\mathbb{P}_0$. Importantly, there are not any parametric restrictions on how the outcome vector $\mathbf{Y}$ is generated; instead, it may be any function of the covariates $(E, \mathbf{W})$, the intervention $A$, and unmeasured factors $U_{\mathbf{Y}}$.  Altogether, our statistical model is semi-parametric.}

{As with the counterfactual outcome vector $\mathbf{Y}(a)$}, we can summarize the observed outcome vector $\mathbf{Y}$ in a variety of ways. 
We again consider weighted sums of the individual-level outcomes within a cluster: 
\begin{align}\label{eq:cluster_y}
    Y_j^c\equiv \sum_{i=1}^{N_j}\alpha_{ij}Y_{ij}
\end{align} 
where $\alpha_{ij}$ matches the definition of the cluster-level counterfactual outcome in Eq.~\ref{Eq:ClusterCF}.
For the remainder of the paper, we focus the cluster-level outcome $Y^c_j$ defined as the empirical mean  within each cluster, corresponding to weights $\alpha_{ij}=\frac{1}{N_j}$ for $i=\{1,\ldots,N_j\}$. However, as previously discussed, we can consider alternative weighting schemes $\alpha_{ij}$ depending on our research question.

To identify the expected counterfactual outcome 
as a function of the observed data distribution and define our target statistical estimand, we require the following two assumptions. Both are satisfied by design in a CRT. First, there must be no unmeasured confounding, such that $A \independent \mathbf{Y}(a)$. Second, there must be a positive probability of receiving each treatment-level: $\mathbb{P}_0(A=a)>0$. Given these conditions are met in a CRT, we can express the cluster-level causal parameter $\Phi^c(a)=\mathbb{E}[Y^c(a)]$ as the expected cluster-level outcome under the treatment-level of interest $\mathbb{E}_0[Y^c|A=a]$, where subscript 0 is used to denote the observed data distribution $\mathbb{P}_0$; a proof is provided in \cite{hierarchical}.
Likewise, the individual-level causal parameter $\Phi(a)=\mathbb{E}[Y(a)]$ equals the expected individual-level outcome {under treatment-level of interest: $\mathbb{E}_0[Y | A=a]$,}
 recognizing the slight abuse to notation because the  observed data $O \sim \mathbb{P}_0$ are at the cluster-level.  

We can gain efficiency in CRTs by adjusting for baseline covariates (e.g., \cite{rubin_vdl_max_eff_2008, adaptive_prespec, murray, moultonhayes}).
Specifically, the treatment-specific expectation of the cluster-level outcome $\mathbb{E}_0[Y^c|A=a]$ can be expressed as the conditional expectation of the cluster-level outcome $Y^c$ given the treatment-level of interest $a$ and the baseline covariates ($E,\mathbf{W})$, averaged over the covariate distribution:  $\mathbb{E}_0[\mathbb{E}_0(Y^c|A=a,E,\mathbf{W})]$.
In practice, we cannot directly adjust for the entire matrix of individual-level covariates $\mathbf{W}$ during estimation. However, we can still improve precision by including lower-dimensional summary measures of $\mathbf{W}$ in the adjustment set or by aggregating the individual-level conditional mean outcome to the cluster-level
(details in Appendix~B of the Supplementary Materials) \cite{hierarchical}. For simplicity, we use $W^c$ to denote either approach to including individual-level covariates in our cluster-level statistical estimand, defined as
\begin{align}\label{eq:target_estimand_cluster}
      \Psi^c_0(a)\equiv \mathbb{E}_0[\mathbb{E}_0(Y^c|A=a,E,{W^c})]
\end{align}
As previously discussed, under the randomization and positivity assumptions, both holding by design, this equals the treatment-specific mean of the cluster-level counterfactual outcomes $\Phi^c(a)=\mathbb{E}[Y^c(a)]$.
  
Likewise, our statistical estimand  for the treatment-specific mean of the individual-level counterfactual outcomes $\Phi(a)=\mathbb{E}[Y(a)]$ is 
  \begin{align}\label{eq:target_estimand_indv}
\Psi_0(a)\equiv \mathbb{E}_0[\mathbb{E}_0(Y|A=a,E,W)]     
  \end{align}
 again acknowledging the slight abuse to notation, because subscript 0 denotes the  distribution of the cluster-level data $O\sim \mathbb{P}_0$.
Our individual-level estimand $\Psi_0(a)$ is the conditional expectation of the individual-level outcome $Y$ given the treatment-level of interest $a$, the cluster-level covariates $E$ and the individual-specific covariates $W$,  averaged over the covariate distribution.  
 For both estimands (Eqs.~\ref{eq:target_estimand_cluster} and~\ref{eq:target_estimand_indv}), we emphasize that covariate adjustment is being used for efficiency gains only and not to control for confounding {\color{black} or other sources of bias, such as missing outcomes or selection \cite{diaz_missing, fiero, Balzer2021twostage, Leyrat2013, Leyrat2014, Li2022, NugentTB2022}.}

As before, we take contrasts of the cluster-level or individual-level estimands corresponding to our causal effect of interest.
To give context for the methods comparison, we will focus on the relative scale for the remainder of the paper; however, our discussion is equally applicable to other scales (e.g., additive or odds ratio). Specifically, the relative effect is identified at the cluster-level as 
\begin{align}
    \frac{\Psi^c_0(1)}{\Psi^c_0(0)} = \frac{ \mathbb{E}_0[\mathbb{E}_0(Y^c|A=1,E,{W^c})]}{\mathbb{E}_0[\mathbb{E}_0(Y^c|A=0,E,{W^c})]}
\label{Eq:ClusterRR}    
\end{align}
and at the individual-level as 
\begin{align}
        \frac{\Psi_0(1)}{\Psi_0(0)} = \frac{ \mathbb{E}_0[\mathbb{E}_0(Y|A=1,E,W)]  }{\mathbb{E}_0[\mathbb{E}_0(Y|A=0,E,W)]  }
\label{Eq:IndvRR}
\end{align}
As detailed below, most analytic methods 
naturally only estimate one of the above statistical estimands, 
whereas few have the flexibility to estimate both.
While  we focused on identification of population-level effects, the extensions to sample effects are fairly straightforward, as discussed in  \cite{balzer_sate}.

\section{Statistical Estimation and Inference} \label{subsec:statistical-summaries}

In this section, we compare the methods commonly used to analyze CRTs. 
We describe their target of inference and their ability to  adjust for baseline covariates to improve precision and thereby improve statistical power. 
We broadly consider two classes of estimation methods: approaches using only cluster-level data and approaches using both individual- and cluster-level data. 
The former immediately aggregate the data to the cluster-level and can only adjust for cluster-level covariates, while the latter allow for adjustment of individual-level covariates, an appealing option, as these pair naturally with individual-level outcomes.
Examples of cluster-level approaches include the t-test
and cluster-level TMLE. Cluster-level approaches naturally target cluster-level effects (e.g., Eq.~\ref{Eq:ClusterRR}), but with the appropriate choice of weights can also target individual-level effects (e.g., Eq.~\ref{Eq:IndvRR}).
Examples of approaches using individual-level data include GEE, CARE, and Hierarchical TMLE {\cite{moultonhayes, liang-zeger, hierarchical}}. 
These individual-level approaches often estimate different causal effects, as detailed below.
 To the best of our understanding, {\color{black}of the algorithms discussed here,}
 Hierarchical TMLE is the only individual-level approach that can estimate effects defined at the cluster-level  (e.g., Eq.~\ref{Eq:ClusterRR}) or at the individual-level (e.g.,  Eq.~\ref{Eq:IndvRR}). {\color{black}It should be possible to modify G-computation and inverse probability weighting to target both cluster-level and individual-level effects, but these extensions are beyond the scope of this paper.} {\color{black} It is not well understood how to incorporate weights for GEE \cite{Wang2021weightGEE}.}

 We now define the notation used throughout this section. Recall the cluster level-outcome $Y^c$ is defined as a weighted sum of individual-level outcomes, as in Eq.~\ref{eq:cluster_y}. 
 We denote the conditional expectation of the cluster-level outcome $Y^c$ given the cluster-level intervention $A=a$ and covariates $(E, W^c)$ as
  \begin{align}
     \mu^c(a,E,{W}^c)&\equiv \mathbb{E}(Y^c|A=a, E, {W}^c)
      \label{eq:outcome_reg_clust}
\end{align}
Likewise, we denote the conditional expectation of the individual-level outcome $Y$ given the  cluster-level intervention $A=a$, the cluster-level covariates $E$, and that individual's covariates $W$ as
\begin{align}
     \mu(a, E,W)&\equiv \mathbb{E}(Y|A=a, E, W)
     \label{eq:outcome_reg_indv}
 \end{align}
Throughout, we refer to Eq.~\ref{eq:outcome_reg_clust} and to Eq.~\ref{eq:outcome_reg_indv} as the cluster-level and individual-level outcome regressions, respectively. The unadjusted expectations of the cluster-level and individual-level outcomes within treatment arm $a$ are defined as $\mu^c(a)\equiv \mathbb{E}(Y^c|A=a)$ and $\mu(a)\equiv \mathbb{E}(Y|A=a)$, respectively.
  We denote the cluster-level propensity score as 
  \begin{align}\label{eq:clust_prop_score}
    \pi^c(a|E,{W}^c)\equiv \mathbb{P}(A=a|E,{W}^c)
  \end{align}
and the individual-level propensity score as 
  \begin{align}\label{eq:indv_prop_score}
    \pi(a|E,W)\equiv \mathbb{P}(A=a|E,W)
  \end{align}
 We define unadjusted probabilities $\pi^c(a)$ and $\pi(a)$, analogously.

\subsection{Analytic Approaches Using Cluster-Level Data} \label{section:cluster-level-estimators}

 Cluster-level approaches obtain point estimates and inference after the  individual-level data have been aggregated to the cluster-level \cite{moultonhayes, murray}. Most commonly, this aggregation is done by taking the empirical mean within each cluster. However, as previously detailed in \Cref{Section:effects}, we can consider several ways to summarize the individual-level data to the cluster-level (i.e., different $\alpha_{ij}$ weighting schemes).  

\subsubsection{Unadjusted Effect Estimator}
\label{Sec:Unadjusted}
 Once the data are  aggregated to the cluster-level, a common approach for  estimation and inference is based on contrasts of the treatment-specific average outcomes:
  \begin{align}
      \hat{\mu}^c(a)=\frac{1}{J}\sum_{j=1}^{J} \frac{\mathbbm{1}(A_j=a)}{\hat{\pi}^c(a)}Y_j^c 
\label{Eq:unadj}
  \end{align} 
  where $\hat{\pi^c}(a)$ denotes the unadjusted estimate of the cluster-level propensity score (i.e., the proportion of clusters in the trial receiving treatment-level $A=a$).
  For simplicity for the remainder of the manuscript,  we assume the trial has equal allocation of arms, such that $\hat{\pi}^c(a)=1/2$; however, our results should generalize to trials with more than 2 arms and to imbalanced trials. Then, if we let $Y^c_{a,k}$ denote the cluster-level outcome for observation $k=\{1,\ldots, J/2\}$ in treatment arm $A=a$, the treatment-specific mean simplifies to 
  $\hat{\mu}^c(a)= \frac{1}{J/2}\sum_{k=1}^{J/2} Y_{a,k}^c$.
  In PTBi, $\hat{\mu}^c(a)$ represents the average incidence of 28-day infant mortality among facilities  that received treatment-level $A=a$.   
We obtain a point estimate of the cluster-level effect by contrasting  $\hat{\mu}^c(1)$ and $\hat{\mu}^c(0)$ on the scale of interest and obtain statistical inference using the t-distribution. Suppose, for example, we were interested in the  cluster-level average treatment effect $\mathbb{E}[Y^c(1)] - \mathbb{E}[Y^c(0)]$; then our point estimate would be $\hat{\mu}^c(1)-\hat{\mu}^c(0)$ and we would test the null hypothesis using a Student's t-test.
Statistical power may be improved by considering alternative weighting schemes when summarizing individual-level outcomes to the cluster-level \cite{moultonhayes};  however, as previously discussed,  different weights $\alpha_{ij}$ 
imply different target effects. 

 For the relative effect, applying the logarithmic transformation is sometimes recommended when the cluster-level summaries are skewed,  which may be more common for rate-type outcomes \cite{moultonhayes}. 
 However, it is important to note that 
 depending on how this transformation is implemented, estimation and inference may be for the ratio of the geometric means, as opposed to the ratio of the arithmetic means. (Recall for  $J$ observations of some variable $X$, the geometric mean is $\left(\prod_{i=1}^J X_i \right)^{1/J}$, whereas the arithmetic mean is $1/J \sum_{i=1}^J X_i$).  
 Specifically, suppose we first take the log of the cluster-level outcomes and then take the average within each arm:
\begin{align}
\bar{l}_a \equiv \frac{1}{J/2}\sum_{k=1}^{J/2} log (Y^c_{a,k})
      =log \left( \prod_{k=1}^{J/2} Y_{a,k}^{c} \right)^{\frac{1}{J/2}} 
\end{align}
where again $Y^c_{a,k}$ denotes the cluster-level outcome for cluster $k=\{1,\ldots, J/2\}$ in treatment arm $a$.
Applying a t-test to the difference in these treatment-specific means $\bar{l}_1 - \bar{l}_0$ (and then exponentiating) targets the ratio of the geometric means.
{\color{black}Continuing our toy example from Section~\ref{Section:effects}, the arithmetic mean of the cluster-level outcomes in the control was $\Phi^{c,J}(0)=0.31$, while the geometric mean of the cluster-level control outcomes would be 0.26.}
To avoid changing the target of inference, we can instead apply the Delta Method to obtain point estimates and inference for the ratio of arithmetic means, as in Eq.~\ref{Eq:ClusterRR}. We refer the reader to \cite{Moore2009} for more details.

 \subsubsection{Cluster-Level TMLE with Adaptive Prespecification} 
   \label{section:cluster-tmle-adapt}

 As previously discussed, statistical power is often improved by adjusting for baseline covariates that are predictive of the outcome.
Once the data have been aggregated to the cluster-level,  we can proceed with estimation and inference for {\color{black} the cluster-level estimand }
   $\Psi_0^c(a)=\mathbb{E}_0[\mathbb{E}_0(Y^c|A=a,E, W^c)]$, using methods for i.i.d. data.
   Examples of common algorithms for $\Psi_0^c(a)$ include 
   parametric G-computation, inverse probability of treatment weighting estimators (IPTW), and TMLE \cite{rose_tl}.
  Due to treatment randomization, these algorithms will be consistent, even under misspecification of the outcome regression \cite{rct_rosenblum_2010}. 
Given that G-computation and IPTW  are well-established approaches, we focus on TMLE, which is a general class of double robust, semiparametric efficient, plug-in estimators \cite{rose_tl}.
  Here, we briefly review the steps of a cluster-level TMLE and then present a solution for 
 optimal  selection of the adjustment covariates
   in trials with limited numbers of clusters. {\color{black}We conclude with a discussion of how to apply weights to the cluster-level TMLE to estimate an individual-level estimand (e.g., $\Psi_0(a)=\mathbb{E}_0[\mathbb{E}_0(Y|A=a, E, W)]$).}
  In the next section,  we discuss an alternative implementation of TMLE that can harness both individual-level and cluster-level covariates to increase precision, while maintaining Type-I error control.
 
To implement a cluster-level TMLE {\color{black} of the cluster-level effect}, we first obtain an initial estimator of the expected cluster-level outcome $\mu^c(A,E,W^c)$.
Next, we update this  initial estimator  $\hat{\mu}^c(A,E,{W}^c)$ using information contained in the estimated propensity score $\hat{\pi^c}(a|E,W^c)$.  Specifically, we define the ``clever covariate'' as the inverse of the estimated propensity score for cluster $j$: \[
\hat{H}^c(a,E_j, W_j^c)=\frac{\mathbbm{1}(A_j=a)}{\hat{\pi^c}(a|E_j,W^c_j)}
\] 
Then on the logit-scale, we regress the cluster-level outcome $Y^c_j$  on the covariates  $\hat{H}^c(1,E_j, W^c_j)$ and $\hat{H}^c(0,E_j, W^c_j)$ with the initial estimator $\hat{\mu}^c(A_j, E_j, W^c_j)$ as the offset. 
This provides the following targeted estimator, while simultaneously solving the efficient score equation: 
\begin{align}\label{eq:update_step}
\hat{\mu}^{c*}(a, E, W^c) =logit^{-1}\big[logit(\hat{\mu}^c(a,E,W^c))+\hat{\epsilon}_1\hat{H}^c(1,E,W^c) + \hat{\epsilon}_0\hat{H}^c(0,E,W^c)\big]
\end{align}
where $\hat{\epsilon}_1$ and $\hat{\epsilon}_0$ denote the estimated coefficients for $\hat{H}^c(1,E,W^c)$ and $\hat{H}^c(0,E,W^c)$, respectively. Finally, we obtain a point estimate of the treatment-specific mean $\Psi_0^c(a)$ by averaging the targeted predictions of the cluster-level outcomes across the $J$ clusters:
 $$\hat{\Psi}^{c*}(a)=\frac{1}{J}\sum_{j=1}^{J} \hat{\mu}^{c*}(a,E_j,{W}^c_j)$$

To  evaluate the intervention effect, we contrast our  estimates $\hat{\Psi}^{c*}(1)$ and  $\hat{\Psi}^{c*}(0)$ on the scale of interest and apply the Delta Method for inference. 
The variance of asymptotically linear estimators, such as the TMLE, may be estimated using the estimator's influence function \cite{rose_tl}. 
These types of estimators enjoy properties that follow from the Central Limit Theorem, allowing us to construct 95\% Wald-type confidence intervals. As a finite sample approximation to the normal distribution, we recommend using the t-distribution with $J-2$ degrees of freedom.   
For the treatment-specific mean $\Psi^{c}_0(a)$, for example, the influence function and asymptotic variance for the cluster-level TMLE are well-approximated as
$\hat{D}^c(a)=\hat{H}^c(a,E,W^c)\times \left(Y^c - \hat{\mu}^{c*}(a,E,W^c) \right) + \hat{\mu}^{c*}(a,E,W^c) - \hat{\Psi}^{c*}(a)$ and $\hat{Var}[\hat{D}^c(a)]/J$, respectively.

In a CRT, the propensity score $\pi^c(a,E,W^c)=\pi^c(a)$ is known and does not need to be estimated. 
However, further gains in efficiency can be achieved through estimation of the propensity score \cite{rct_rosenblum_2010, rubin_vdl_max_eff_2008}. If both the cluster-level outcome regression and cluster-level propensity score are consistently estimated, 
the TMLE will be an asymptotically efficient estimator. However, consistent estimation of the outcome regression is nearly impossible when using an \emph{a priori}-specified regression model. 

To improve precision while preserving Type-I error control, we previously proposed \emph{``Adaptive Prespecification''}, a supervised learning approach using sample-splitting to choose the adjustment set that maximizes efficiency \cite{adaptive_prespec}. 
In brief, we prespecify a set of candidate generalized linear models (GLMs) for the cluster-level outcome regression $\mu^c(a,E,W^c)$ and propensity score $\pi^c(a|E,W^c)$. {\color{black}To avoid forced adjustment at the detriment of precision, the unadjusted estimator should always be included as a candidate.}
We also prespecify a cross-validation scheme; for small trials (e.g, $J \leq 30$), we recommend leave-one-cluster-out. To measure performance, we prespecify the squared influence function as our loss function. Then we choose the candidate estimator of $\mu^c(a,E,W^c)$ that minimizes the cross-validated variance estimate using the influence function based on the known propensity score (i.e., $\pi^c(a)=0.5$).  We then select the candidate estimator of propensity score $\pi^c(a |E,W^c)$ that further minimizes the cross-validated variance estimate using the influence function \textit{when} combined with the previously selected estimator $\hat{\mu}^c(a,E,W^c)$. 
Together, the selected estimators $\hat{\mu}^c(a,E, W^c)$ and $\hat{\pi}^c(a|E,W^c)$ form the ``optimal'' TMLE
according to the principle of empirical efficiency maximization \cite{rubin_vdl_max_eff_2008}.

{\color{black}Application of TMLE with Adaptive Prespecification to real  data from the SEARCH CRT resulted in notable precision gains  \cite{Havlir2019, Balzer2021twostage}. As compared to the unadjusted estimator, TMLE was 4.6-times more efficient for the effect on HIV incidence, 2.6-times more efficient for the effect on the incidence tuberculosis, and 1.8-times more efficient for the effect on hypertension control. Given that this cluster-level TMLE adjusted for at most 2 cluster-level variables, these gains in efficiency may seem surprising. However, we  believe that they demonstrate the power of using Adaptive Prespecification to flexibly select the optimal adjustment strategy to maximize empirical efficiency. These gains are also seen in a recent extension of Adaptive Prespecification for flexible adjustment of many covariates in trials with a larger number of randomized units \cite{Balzer2022FancyAPS}.}

\subsubsection{Targeting an Individual-level Effect with an Estimator Using Cluster-level Data}

With or without Adaptive Prespecification, the cluster-level TMLE naturally estimates cluster-level parameters (e.g., Eq.~\ref{Eq:ClusterRR}) corresponding to contrasts of cluster-level counterfactuals, such as $\mathbb{E}[Y^c(1)] \div \mathbb{E}[Y^c(0)]$. However, the cluster-level TMLE can also estimate individual-level parameters (e.g., Eq.~\ref{Eq:IndvRR}) corresponding to contrasts of individual-level counterfactuals, such as  $\mathbb{E}[Y(1)] \div \mathbb{E}[Y(0)]$. To do so, we include weights $J/N_T \times 1/\alpha_{ij}$ in each step of the cluster-level analysis {\color{black}(derivation and \texttt{R} code in the Supplementary Materials)}. This approach may be relevant when data are only available at the cluster-level, but interest is in an individual-level effect.
Finally, we note that the unadjusted effect estimator can be considered a special case of the TMLE where the adjustment set is empty: $(E,W^c)=\{\emptyset\}$. Therefore, the unadjusted effect estimator based on cluster-level data can also estimate an individual-level effect if the appropriate weights are applied.

\subsection{Analytic Approaches Using  Individual-Level Data}

We now discuss how to leverage individual-level covariates when estimating effects in CRTs. This is done  by aggregating to the cluster-level \textit{after} estimating the expected individual-level outcome or by
implementing a fully individual-level approach. In all cases,  clustering must be accounted for during variance estimation. 
Estimators using individual-level data vary in their flexibility to estimate both cluster-level effects (e.g., Eq.~\ref{Eq:ClusterRR}) and individual-level effects (e.g., Eq.~\ref{Eq:IndvRR}).

\subsubsection{Hierarchical TMLE} \label{section:hier_tmle}

In \Cref{section:cluster-level-estimators}, we discussed a cluster-level TMLE for estimating effects in CRTs based on  aggregating the data to the cluster-level. An alternative and equally valid approach for CRT analysis is to 
ignore clustering when obtaining a point estimate
and then account for clustering during variance estimation and statistical inference \cite{ware-laird, Schnitzer2014}. {\color{black}Such an approach naturally estimates an individual-level parameter, such as $\Psi_0(a)=\mathbb{E}_0[\mathbb{E}_0(Y|A=a,E,W)]$.}
We now present an individual-level TMLE using this alternative approach and refer to it as ``Hierarchical TMLE'' to emphasize its distinction from the standard TMLE for an individually randomized trial with $N_T$ i.i.d. participants.
In Appendix B, we also present a ``Hybrid TMLE'' that  obtains an initial estimate of the cluster-level outcome regression based on aggregates of the individual-level outcome regression (i.e., $\hat{\mu}^c(A_j,E_j,W^c_j)= \sum_i^{N_j} \alpha_{ij} \hat{\mu}(A_j,E_j,W_{ij})$)  and then proceeds with estimation and inference as outlined  in Section~\ref{section:cluster-tmle-adapt}.
Both approaches leverage the natural pairing of individual-level outcomes with individual-level covariates, extend to a pair-matched design \cite{hierarchical}{\color{black}, and can be combined with weights to estimate effects at the cluster-level or individual-level.}

To implement Hierarchical TMLE {\color{black} for the individual-level effect}, we pool participant-level data across clusters to obtain  estimators of the individual-level outcome regression  $\mu(A, E, W)$ and the individual-level propensity score $\pi(a|E,W)$. The initial outcome regression estimator $\hat{\mu}(A,E,W)$ is then updated based on the estimated propensity score $\hat{\pi}(a|E,W)$. 
As before, we calculate the ``clever covariate'', but now at the individual-level: \[
\hat{H}(a, E_j, W_{ij})=\frac{\mathbbm{1}(A_j=a)}{\hat{\pi}(a|E_j, W_{ij})}
\]
for $a=\{1,0\}$. Then on the logit-scale, we regress the individual-level outcome $Y_{ij}$ on the individual-level covariates $\hat{H}(1,E_j, W_{ij})$ and $\hat{H}(0, E_j, W_{ij})$ with the initial individual-level estimator $\hat{\mu}(A_j,E_j,W_{ij})$ as the offset.
This provides the following updated estimator of the expectation of the individual-level outcome, while simultaneously solving the efficient score equation: 
\begin{align}\label{eq:update_step2}
\hat{\mu}^{*}(a, E, W) =logit^{-1}[logit(\hat{\mu}(a,E,W))+\hat{\epsilon}_1\hat{H}(1, E,W) + \hat{\epsilon}_0\hat{H}(0,E,W)]
\end{align}
where $\hat{\epsilon}_1$ and $\hat{\epsilon}_0$ now denote the estimated coefficients for $\hat{H}(1,E,W)$ and $\hat{H}(0,E,W)$.
Then we  obtain a point estimate of the treatment-specific mean $\Psi_0(a)$ by averaging these targeted predictions:  $$\hat{\Psi}^*(a)=
\frac{1}{N_T}\sum_{j=1}^J  \sum_{i=1}^{N_j} \hat{\mu}^{*}(a,E_j,W_{ij})$$
where, again, $N_T$ denotes the total number of participants across all clusters. 
{\color{black}Thus, this implementation of Hierarchical TMLE to estimate an individual-level parameter is nearly identical to standard TMLE for i.i.d. data. Key distinctions are in sample-splitting (if used) and variance estimation, both of which must respect the cluster as the independent unit in CRTs.} 
Specifically, to estimate the variance of Hierarchical TMLE for $\hat{\Psi}^*(a)$,  we aggregate an individual-level influence function to the cluster-level and then take the sample variance of the estimated cluster-level influence function, scaled by the number of independent units $J$. 
For example, the influence function for this TMLE of $\hat{\Psi}^*(a)$ is well-approximated by 
 $\hat{D}_j(a)= \sum_{i=1}^{N_j} J/N_T \times \hat{D}_{ij}(a)$
 where $\hat{D}_{ij}(a)=
 \hat{H}(a,E_{j},W_{ij})\times {\color{black}\left(Y_{ij}- \hat{\mu}^{*}(a,E_{j},W_{ij}) \right)} + \hat{\mu}^{*}(a,E_{j},W_{ij}) - \hat{\Psi}^{*}(a)$. {\color{black}(See Schnitzer et al. for a proof \cite{Schnitzer2014}.)  
 Altogether, this implementation of Hierarchical TMLE naturally estimates individual-level effects (e.g., Eq.~\ref{Eq:IndvRR}) and is analogous to using an independent working correlation matrix with the robust variance estimator in Generalized Estimating Equations (GEE) \cite{Wang2021weightGEE}, described below.

 Unlike GEE, however, Hierarchical TMLE can also easily estimate cluster-level effects (e.g., Eq.~\ref{Eq:ClusterRR}) \cite{hierarchical}. To do so, we incorporate the   $\alpha_{ij}$ weights throughout the analysis to obtain cluster-level point estimates and inference. Specifically, when obtaining a point estimate, we  aggregate the targeted predictions within clusters before averaging across clusters: $\hat{\Psi}^{c*}(a)=1/J \sum_j^J \sum_i^{N_j}\alpha_{ij} \hat{\mu}^*(a, E_j, W_{ij})$. Likewise, we estimate a cluster-level influence function for this TMLE of $\hat{\Psi}^{c*}(a)$ as $\hat{D}_j^c(a)= \sum_{i=1}^{N_j} \alpha_{ij} \times \hat{D}_{ij}(a)$
 where $\hat{D}_{ij}(a)=
 \hat{H}(a,E_{j},W_{ij})\times{\color{black}\left(Y_{ij}- \hat{\mu}^{*}(a,E_{j},W_{ij}) \right)}+ \hat{\mu}^{*}(a,E_{j},W_{ij}) - \hat{\Psi}^{c*}(a)$.
 Further details are available in Balzer et al. \cite{hierarchical} and \texttt{R} code is provided in the Supplementary Materials.

 For either the cluster-level or individual-level effect, the variance of Hierarchical TMLE is well-approximated by the variance of the cluster-level influence function, scaled by the number of independent units $J$.}
 As before, the Delta Method is applied to obtain point estimates and inference for the intervention effect on any scale of interest. Again, we recommend the t-distribution with $J-2$ degrees of freedom for confidence interval construction and testing the null hypothesis.  
Hierarchical TMLE will be an asymptotically efficient estimator if the outcome regression and  propensity score are consistently estimated at reasonable rates.
In practice, we again recommend using Adaptive Prespecification to select the optimal adjustment strategy to maximize efficiency.

\subsubsection{Covariate-Adjusted Residuals Estimator (CARE)}
\label{subsection:CARE}

The covariate-adjusted residuals estimator (CARE) was first proposed in Gail et al. \cite{gail-care} and later popularized by Hayes and Moulton \cite{moultonhayes}.
CARE is implemented by pooling individual-level data across clusters and then regressing the individual-level outcome $Y$ on the individual-level and cluster-level covariates of interest $(W,E)$, but not the cluster-level intervention $A$. Then the predictions from this regression are aggregated to the cluster-level. Finally, a t-test comparing the mean residuals (i.e., the discrepancies between observed and predicted outcomes) by arms is performed, since the average residuals should be the same between arms  under the null hypothesis.
CARE naturally targets cluster-level effects, as in  Eq.~\ref{Eq:ClusterRR}{\color{black}, and it is not immediately obvious how to incorporate weights to target individual-level effects, as in Eq.~\ref{Eq:IndvRR}.}

As a concrete example, suppose our goal is to estimate the cluster-level effect on the relative scale and the individual-level outcome $Y$ is binary. To implement CARE in this setting, we first fit an individual-level logistic regression, such as
\begin{align}
  \mathbb{E}(Y|E, W)= logit^{-1}[ \alpha + \beta_E E + \beta_W W ] 
\end{align}
where $\beta_{E}$ and $\beta_{W}$ denote the magnitude by which the log odds of the outcome for the $i$th individual in the $j$th cluster is affected (linearly) by the cluster-level covariates $E_j$ and individual-level covariates $W_{ij}$, respectively. From this regression we obtain the expected number of events in the $j$th cluster  as $e_{j}=\sum_{i}^{N_j} {\color{black}logit^{-1}}(\hat{\alpha} + \hat{\beta_E}E_j + \hat{\beta_W}W_{ij})$ and compare it with  the observed number of events $d_{j}= \sum_i^{N_j}Y_{ij}$
through 
 ratio-residuals:
 $$R_{j}=\frac{d_{j}}{e_{j}} $$ 
Hayes and Moulton \cite{moultonhayes} note that these ratio-residuals are often right-skewed and recommend a logarithmic transformation. 
 Specifically, they recommend applying a t-test to obtain point estimates and inference for the difference in the treatment-specific averages of the log-transformed residuals: 
\begin{align}
 \frac{1}{J/2}\sum_{k=1}^{J/2} log (R_{1,k}) -  \frac{1}{J/2}\sum_{k=1}^{J/2} log (R_{0,k})  
      =log \left( \prod_{k=1}^{J/2} R_{1,k} \right)^{\frac{1}{J/2}} - log \left( \prod_{k=1}^{J/2} R_{0,k}\right)^{\frac{1}{J/2}} 
\end{align}
where $R_{a,k}$ denotes the ratio-residual for cluster $k=\{1,\ldots, J/2\}$ in arm $a$. 
 As detailed in Section~\ref{Sec:Unadjusted}, after exponentiation, we recover estimates and inference for the ratio of the geometric means and thereby a different causal effect than the standard risk ratio, given in Eq.~\ref{Eq:ClusterRR}. 
 A straightforward extension to pair-matched design is illustrated in \cite{moultonhayes}.

\subsubsection{Generalized Estimating Equations (GEE)}
\label{section: gee}

We now consider a class of estimating equations, sometimes referred to as ``population-average models,'' for estimating effects in CRTs \cite{ware-laird}. 
In GEE, estimation and inference is conducted at the individual-level and a working correlation matrix is used to account for the dependence of outcomes within clusters. GEE naturally targets individual-level effects, as in Eq.~\ref{Eq:IndvRR}. 
{\color{black} As with CARE, it is unclear whether weights can be incorporated to instead target cluster-level effects, as in Eq.~\ref{Eq:ClusterRR}. Indeed, recent work by Wang et al. \cite{Wang2021weightGEE} suggest that when targeting the cluster-level effect, inappropriate use of weights in GEE can result in meaningful bias.}

In GEE, the expected individual-level outcome is modeled a function of the treatment and possibly covariates of interest 
\cite{ware-laird, hubbard_ahern}. 
Specifically, consider the following ``marginal model'' for the expected individual-level outcome $\mathbb{E}(Y|A)$: 
\begin{align}\label{eq:marg_model}
\mu(A_j) = g^{-1}(\beta_0+\beta_A A_j)
\end{align}
where $g^{-1}(\cdot)$ denotes the inverse-link function.  Commonly, the identity link is used for continuous outcomes, the log-link for count outcomes, and the logit-link for binary outcomes. 
Effect estimation in GEE is usually done by obtaining a point estimate and inference for the treatment coefficient $\beta_A$. 
{Then at the individual-level, $\beta_A$ represents the additive causal effect for the identity link; $e^{\beta_A}$ represents the relative effect for the log-link, and $e^{\beta_A}$ represents the  odds ratio effect for the logit-link.}
In other words, the link function often determines the scale on which the effect is estimated.

As with other CRT approaches, GEE may improve efficiency by adjusting for covariates. 
Consider, for example, the following ``conditional model'' for the expected individual-level outcome $\mathbb{E}(Y|A,E,W)$: 
\begin{align}\label{eq:adj_gee}
   \mu(A_j, E_j, W_{ij})= g^{-1}(\beta_0+\beta_A A_j +  \beta_E E_j + \beta_W W_{ij})
\end{align}
where again $g^{-1}(\cdot)$ denotes the inverse-link function. 
Except for linear and log-linear models without interaction terms, the interpretation of $\beta_A$ is generally not the same as in the marginal model \cite{hubbard_ahern} For the logistic link function, for example, $\beta_A$ in Eq.~\ref{eq:adj_gee} would yield the conditional log-odds ratio, instead of the marginal log-odds ratio. However, a recent modification to GEE, presented in the next subsection, allows for estimation of marginal effects, while adjusting for individual-level or cluster-level covariates.

For either a marginal or conditional specification, the GEE estimator solves the following equation:
$$ \sum_{j=1}^J \mathbf{D}_j^T \mathbf{V}_j^{-1}(\mathbf{Y}_j-\bm{\mu}_j)=0$$
 where $\bm{\mu}_j$ is the vector containing individual-level outcome regressions for cluster $j$,  $\mathbf{D}_j=\frac{\delta \bm{\mu}_j}{\delta \beta}$ is the gradient matrix, and $\mathbf{V}_j$ is the
 working correlation matrix used to account for dependence of individuals within a cluster $j$
  \cite{ware-laird}.
GEE yields a consistent point estimate of $\beta_A$ under the marginal model, even if the correlation matrix has been misspecified. 
It is worth noting, however, that the interpretation of the estimated effect changes  subtly when using alternative correlation matrices \cite{hubbard_ahern}. Additionally, under misspecification of the correlation matrix, the usual standard errors obtained are not valid, and the sandwich variance estimator must be used \cite{ware-laird}. In general, unless the number of clusters is relatively large and the number of participants within cluster is relatively small, the sandwich-based standard errors can underestimate the true variance of $\beta$ and  yield confidence intervals with coverage probability below desired nominal level. Fortunately, several corrections to variance estimation exist 
\cite{murray, gee_correction, Fay2001}.
Finally, for a CRT where the intervention is randomized within matched pairs of clusters, a pair-matched analysis can conducted by specifying a fixed effect for the pair, while maintaining the correlation structure at the cluster-level. However, pair-matched analyses are generally discouraged for studies with fewer than 40 clusters \cite{murray, gee_correction, moultonhayes}.

\subsubsection{Augmented-GEE}

As discussed in the previous subsection, the treatment coefficient $\beta_A$ resulting from  GEE  does not always correspond to the marginal effect. Recently, a modification to GEE was proposed to ensure $\beta_A$ can be interpreted as a marginal effect, while simultaneously adjusting for baseline covariates to improve efficiency \cite{aug-gee}. This approach is referred to as  Augmented-GEE (Aug-GEE) and naturally targets individual-level effects, as in  Eq.~\ref{Eq:IndvRR}. As with standard GEE, {\color{black}it is unclear if weights can be incorporated in Aug-GEE to instead target cluster-level effects, as in Eq.~\ref{Eq:ClusterRR}. Additionally, as with standard GEE,} the link function $g(\cdot)$ determines the scale on which the effect is estimated in Aug-GEE (i.e., additive for the identity link, relative for the log-link, and odds ratio for the logit-link).

As commonly implemented, Aug-GEE modifies GEE 
by including an additional ``augmentation" term, which incorporates the conditional expectation of the individual-level outcome $\mu(A_j,E_j,W_{ij})$. 

The general form of the Aug-GEE for a binary treatment is given by 
\begin{align}
\label{Eq:AugGEE}
\sum_{j=1}^J \Big[ \mathbf{D}_j^T \mathbf{V}_j^{-1}(\mathbf{Y}_j-\bm{\mu}_j(A_j))-\sum_{a=0}^1\big[\mathbbm{1}(A_j=a)-\pi^c(a)\big]\gamma_a\Big ] = 0
\end{align}
with augmentation term 
$$\gamma_a=\mathbf{D}_j^T \mathbf{V}_j^{-1}(\bm{\mu}_{j}(a,E_j, W_{ij})-\bm{\mu}_j(a))$$
For cluster $j$,  $\bm{\mu}_j(A_j)$ denotes the vector of marginal regressions (as in Eq.~\ref{eq:marg_model}) and $\bm{\mu}_j(A_j,E_j,W_{ij})$ denotes the vector of conditional regressions (as in Eq.~\ref{eq:adj_gee}). 
The cluster-level propensity score, defined in 
Eq.~\ref{eq:clust_prop_score}, is treated as known  (i.e., $\pi^c(a)=0.5$). 
By solving Eq.~\ref{Eq:AugGEE}, we can obtain point estimates and inference for the marginal effects. 
As with the other estimators considered, if the conditional outcome regression $\mu(A_j,E_j,W_{ij})$ is misspecified, the resulting estimator is asymptotically normal and consistent; however, it is not efficient \cite{aug-gee}. 
 Indeed, the efficiency of Aug-GEE heavily depends on the matrix $\mathbf{D}^{T}\mathbf{V}^{-1}$. Stephens et al. \cite{stephens_aug-gee14} show how to further improve  Aug-GEE by deriving the semiparametric locally efficient estimator, but ultimately  conclude that the  high-dimensional inverse covariance matrix, required in their approach, presents a substantial barrier to any practical gains.

\section{Simulation Studies}

To examine the finite sample properties of the previously discussed CRT estimators and to demonstrate the practical impact of targeting different causal effects, we conducted two simulation studies. Full \texttt{R} code is provided in the Supplementary Materials. 

\subsection{Simulation I}

We simulated a simplified data generating process reflecting the  hierarchical data structure of the PTBi study, which randomized $J=20$ clusters, corresponding to health facilities. For each cluster $j=\{1,\ldots, 20\}$,  we generated cluster-level covariates $E1_j \sim Norm(2, 1)$ and $E2_j \sim Norm(0, 1)$ and cluster size $N_j \sim Norm(150, 80)$ subject to a minimum  of 30 participants.  
For each cluster $j$, we also simulated random variables $U_{E2_j} \sim Unif(-0.2, 1.5)$ and $U_{E2_j} \sim Unif(-0.5,0.5)$ to act as an unmeasured source of dependence within each cluster. Then for participant $i=\{1,\ldots, N_j\}$ in each cluster $j$, we generated 4 individual-level covariates:
$W1_{ij} \sim Norm(2U_{E1_j}, 0.35)$;  
$W2_{ij} \sim Norm(4 U_{E1_j}, 0.9)$; 
$W3_{ij} \sim Norm(U_{E2_j}, 0.5)$, and  
$W4_{ij} \sim Norm(U_{E2_j}, 0.5)$.

To reflect the PTBi study design, we paired clusters on $E2$ using the nonbipartite matching algorithm \cite{nbp_match}.
Within the pair, one cluster was randomized to the intervention arm ($A_j=1$) and the other to the control arm ($A_j=0$). 
Lastly, we generated the individual-level outcomes %
as a function of the intervention $A_j$, the cluster-level and individual-level covariates, and unmeasured factor $U_{Y_{ij}} \sim Unif(0,1)$: 
    \begin{align}
        Y_{ij} \sim \mathbbm{1}\big[ U_{Y_{ij}} < logit^{-1}(-0.75 - 0.35A_j + 0.8W1_{ij} + 0.4W2_{ij} - 0.3E1_j -0.2A_j W2_{ij}) \big]
    \end{align}
To assess Type-I error control, we generated the outcomes after setting the terms involving the treatment $A$ to 0. For a population of 2500 clusters, we also generated counterfactual outcomes  under the intervention $Y(1)$  and under the control $Y(0)$ by setting $A=1$ and $A=0$, respectively. These counterfactuals were used to calculate the true values of the causal parameters, defined at both the cluster-level and individual-level. Specifically, we calculated the cluster-level relative effect (Eq.~\ref{Eq:ClusterRR}), the individual-level relative effect (Eq.~\ref{Eq:IndvRR}), as well as the geometric incidence ratio (i.e., the ratio of the geometric means of the cluster-level outcomes).

For estimation of the cluster-level relative effect, we implemented  Hierarchical TMLE using Adaptive Prespecification to select from $\{\emptyset, W1, W2, W3, W4\}$ and the cluster-level TMLE with  Adaptive Prespecification to select the optimal adjustment set from $\{\emptyset, W1^c, W2^c, W3^c, W4^c\}$, where $W1^c, \ldots, W4^c$ were the empirical means of their individual-level counterparts.  
For comparison, we also implemented a cluster-level TMLE not adjusting for any covariates, hereafter called the ``unadjusted estimator''.
Inference for the unadjusted estimator and the TMLEs was based on their influence functions.

CARE, GEE, and Aug-GEE relied on a fixed specification of the individual-level outcome regression with main terms adjustment for both cluster-level and individual-level covariates: $\{W1, W2, W3, W4, E1, E2\}$. For CARE, we used the logit-link to obtain outcome predictions in the absence of the treatment, then applied the log-transformation to the ratio residuals, as recommended  by Hayes and Moulton \cite{moultonhayes}, and finally obtained inference with a t-test. For comparison, we also implemented a standard Student's t-test after log-transforming the cluster-level outcomes.  In GEE and Aug-GEE, we used the log-link function
{\color{black} and Fay and Graubard's finite sample variance correction \cite{Fay2001}, as implemented in the \texttt{geesmv}} and \texttt{CRTgeeDR} packages, respectively \cite{geesmv, CRTgeeDR}. 
All algorithms ignored the matched pairs used for intervention randomization.

\subsection{Results of Simulation I}

In these simulations, the cluster-level incidence ratio (Eq.~\ref{Eq:ClusterRR}), targeted by the unadjusted estimator and the TMLEs, was identical to the individual-level risk ratio (Eq.~\ref{Eq:IndvRR}), targeted by GEE and Aug-GEE: 0.83.  The ratio of geometric means, targeted by the t-test and CARE, was 0.81, indicating a slightly larger effect. In other words, the simulated intervention resulted in a relative reduction of the mean cluster-level outcomes of 17\%  on the arithmetic scale and 19\% on the geometric scale. As demonstrated in the second simulation study, there can be substantial divergence between the causal parameters, and it is essential to prespecify a target effect corresponding to the research query.
 
 While the CRT estimators targeted different effects, it is still valid to compare their attained power, defined as the proportion of times the false null hypothesis was rejected at the 5\% significance level,  and confidence interval coverage, defined the proportion of times the 95\% confidence intervals contained the true value of the target effect. Additionally, we examined Type-I error, defined as the proportion of times the true null hypothesis was rejected at the 5\% significance level.
These metrics are shown in \Cref{tab:sim_I_res} for each estimator (for its corresponding target estimand) across 500 iterations of the data generating process, each with $J=20$ clusters.

As shown in Table~\ref{tab:sim_I_res}, the unadjusted estimator of the cluster-level relative effect achieved low statistical power (18\%) with slightly conservative confidence interval coverage (97\%) when there was an effect and controlled  Type-I error  (4\%) under the null. In this simulation, the cluster-level TMLE and Hierarchical TMLE performed similarly and provided substantial efficiency gains over the unadjusted estimator of the same effect. Specifically, the TMLEs achieved a statistical power of 99\% with conservative interval coverage ($\geq$ 95\%) and  strict Type-I error control ($<$5\%) under the null.
{\color{black}Thus, by adaptively adjusting for at most two covariates (one in the outcome regression and one in the propensity score), the TMLEs improved statistical power over the unadjusted estimator by 81\%. In line with previous work \cite{adaptive_prespec, Balzer2021twostage, Benkeser2021}, this re-iterates the importance of a data-driven approach to covariate adjustment that is tailored to maximizing efficiency.}
Specifically, the TMLEs 
differed in their selection of adjustment variables.
For estimation of the outcome regression, the cluster-level TMLE selected $W1^c$ in 54\% of the simulated trials and $W2^c$ in the other 46\%. In contrast, the Hierarchical TMLE selected $W1$ in 82\% of iterations and $W2$ in 18\% of iterations.
The different selections illustrate how the relationship between the cluster-level outcome and cluster-level covariates  can be distinct from the relationship between the individual-level outcome and individual-level covariates --- impacting 
the optimal adjustment strategy. 
Importantly, both approaches avoided adjustment for covariates that were not predictive of the outcome (i.e., $\{W3^c, W4^c\}$ at the cluster-level and $\{W3, W4\}$ at the individual-level). 
For these estimators, additional simulation results, including fewer clusters ($J=10)$, smaller clusters (mean cluster size $N_j$ of 20), 
and for the individual-level effect, are given in the Supplementary Materials. These simulations again demonstrate substantial gains in statistical power from using TMLE, while tightly preserving  confidence interval coverage and Type-I error control under the null. 

{\color{black}The other estimators, relying on fixed specifications of their outcome regressions, did improve power over their unadjusted counterparts, but not to the same extent as the TMLEs. Specifically, for estimation of the geometric incidence ratio, using CARE to adjust for both cluster-level and individual-level covariates improved statistical power to 93\%, vastly surpassing the power of the t-test (14\%). However, CARE failed to maintain nominal confidence interval coverage (86\%).
For the individual-level risk ratio, equal to the cluster-level incidence ratio in these simulations,  GEE offered power improvements over the unadjusted approach (75\% vs. 18\%, respectively). However, GEE exhibited less than nominal confidence interval coverage (89\%) and   inflated Type-I error rates (8\%) under the null. While Aug-GEE, also targeting the individual-level risk ratio, slightly improved power over standard GEE (77\%); it exhibited worse confidence interval coverage (84\%) and  doubled the Type-I error rate (16\%).}

\subsection{Simulation II}

In many CRTs, participant outcomes are influenced by the cluster size. Suppose, for example, that the smallest health facilities have the fewest  resources to the detriment of their patients' health, and largest health facilities are overburdened also to the detriment of their patients' health. In this setting, wide variation in cluster size can result in a divergence between cluster-level and individual-level effects. Intuitively, cluster-level parameters (e.g., Eqs.~\ref{Eq:Meta} and~\ref{Eq:ClusterCausal}) give equal weight to each cluster, regardless of its size, while individual-level parameters (e.g., Eqs.~\ref{Eq:IndvSample} and~\ref{Eq:IndvCausal}) give equal weight to all trial participants. The distinction between the parameters is exacerbated when cluster size interacts with the treatment and is said to be ``informative'' \cite{nevail_ics2014, Seaman2014}. 
Therefore, in the second simulation study, we considered a more  complex data generating process to highlight the distinction between the cluster-level effects and individual-level effects.

We again focused on a setting with $J=20$ clusters, reflecting the PTBi study. For each cluster $j=\{1,\ldots, J\}$, we generated  cluster-level covariates $E1_j \sim Norm(0, 1)$,  $E1_j \sim Norm(0, 1)$, and the cluster size $N_j \sim Norm(400,250)$, again truncated at a minimum of 30 participants. For each participant $i=\{1,\ldots, N_j\}$ in cluster $j$, we generated three individual-level covariates  
$W1_{ij}\sim Norm(U_{E1_j},0.5)$, 
$W2_{ij}\sim Norm(U_{E2_j},0.5)$, and 
$W3_{ij}\sim Norm(U_{E3_j},0.5)$ 
with cluster-specific means $U_{E1_j}\sim Unif(-1,1)$, $U_{E2_j}\sim Unif(-1,1)$,
and $U_{E3_j}\sim Unif(-1,1)$.

 As before, clusters were pair-matched on $E2$, and within each pair, one cluster was randomized to the intervention arm $(A=1)$ and the other to the control arm $(A=0)$. 
 Lastly, we simulated the individual-level outcomes as a function of the treatment, the cluster-level and individual-level covariates, and unmeasured factor $U_{Y_{ij}} \sim Unif(0,1)$: 
\begin{align*}
      Y_{ij} \sim \mathbbm{1}[U_{Y_ij} < logit ^ {-1} 
      (0.5 + W1_{ij}/6 + W2_{ij}/2 + W3_{ij}/4 + E1_j/5 + E2_j/5 - \tilde{N}_j/8 - A\tilde{N}_j/5 )]
\end{align*}
where $\tilde{N}_j= N_j/150$ denotes the scaled cluster size.
Unlike Simulation I, the probability of the individual-level outcome $Y_{ij}$ was a function of cluster size $N_j$. Specifically, the outcome risk was lower for participants in larger clusters, especially large clusters in the intervention arm.
As before, we generated counterfactual outcomes under the intervention and under the control by setting $A=1$ and $A=0$, respectively. Then for a population of 1000 clusters, we calculated the relative effect at the cluster-level (Eq.~\ref{Eq:ClusterRR})
and at the individual-level
(Eq.~\ref{Eq:IndvRR}). 

For this simulation, we focused on the performance of the cluster-level TMLE and Hierarchical TMLE, given their flexibility to estimate a variety of effects and their ability to incorporate baseline covariates to improve precision and statistical power, while maintaining Type-I error control. 
As previously discussed, through the application of weights, the cluster-level TMLE can estimate individual-level effects. Likewise,  Hierarchical TMLE, an  estimation approach based on individual-level data, can estimate cluster-level effects. It is possible that the other CRT estimators have this flexibility, but the needed extensions remain to be fully studied \cite{Wang2021weightGEE}.


In this simulation, each TMLE used Adaptive Prespecification to choose at most two covariates for adjustment if their inclusion improved efficiency as compared to an unadjusted effect estimator. The candidate adjustment set for the cluster-level TMLE included 
$\{\emptyset, W_1^c, W_2^c\}$,
while the candidate adjustment set for individual-level TMLE included  $\{\emptyset, W_1, W_2\}$. 
%
For comparison, we also considered 
a cluster-level TMLE with fixed adjustment for $ W^c_1$ {in the cluster-level outcome regression} and a Hierarchical TMLE with fixed adjustment for $W_1$ {in the individual-level outcome regression}.
Inference was based on an estimate of the influence function. We again focus on the analysis breaking the matches used for randomization.

\subsection{Results of Simulation II }

In these simulations, the true value of the cluster-level relative effect (Eq.~\ref{Eq:ClusterRR}) was  0.78, substantially smaller than the true value of the individual-level relative effect (Eq.~\ref{Eq:IndvRR}) of 0.69. In other words, the intervention resulted in a 22\% relative reduction in the incidence of the outcome and  a 31\% relative reduction in the individual-level risk of the outcome. Of course, this is just one simulation study, and in practice, there is no guarantee that the cluster-level effect will be smaller, or even different, from the individual-level effect.

For this simulation study, \Cref{tab:ics_results} shows the performance of the cluster-level TMLE and Hierarchical TMLE with fixed and adaptive adjustment. Given the differing magnitude of the effects, it is unsurprising that estimators of the individual-level effect, shown on the right, achieved notably higher power than estimators of the cluster-level effect, shown on the left. Specifically, estimators of the cluster-level effect achieved a maximum power of 44\%, while estimators of the individual-level effect achieved a maximum power of 68\%.

Focusing on estimators of the cluster-level effect (\Cref{tab:ics_results}-Left),  all approaches resulted in low  bias,  similar variability ($\sigma$), and good confidence interval coverage ($\geq$96\%). For a given adjustment strategy (fixed or adaptive), there was little practical difference in performance between analyses using cluster-level or individual-level data. {\color{black} However, as expected by theory \cite{adaptive_prespec},} the TMLEs using Adaptive Prespecification achieved higher power ($\approx$44\%) than the TMLEs relying on fixed specification of the outcome regression (40\%).
 For the individual-level effect (\Cref{tab:ics_results}-Right), the gains in power with adaptive adjustment were more substantial. Specifically, TMLEs with fixed adjustment achieved a maximum power of 60\%, while the TMLEs using Adaptive Prespecification achieved a maximum power of 68\%. While the approaches were slightly biased towards the null, all  had good  confidence interval coverage (94\%-96\%). Again we see little practical difference between approaches relying on cluster-level data or utilizing individual-level data. 
As shown in the Supplementary Materials, all approaches also maintained strong Type-I error control under the null.

Altogether the results of this simulation  demonstrate the ability of TMLE to estimate both cluster-level and individual-level effects, while adaptively adjusting for baseline covariates to maximize efficiency. These results also highlight the critical importance of pre-specifying the primary effect measure and using a CRT estimator of that effect. 
Notable bias and misleading inference may arise when the statistical estimation approach is mismatched with the desired causal effect. 
In all settings, the research question should drive the specification of the target effect and thereby the statistical estimation approach \cite{Petersen2014roadmap, ICHE9, BalzerMLcomm2021, Kahan2021}.

\section{Real Data Application: The PTBi Study in Kenya and Uganda} 
\label{section: Real Data}

In East Africa, preterm birth remains a leading risk factor for perinatal mortality, defined as stillbirth and first-week deaths \cite{ku_protocol}. Evidence-based practices, such as use of antenatal corticosteroids and skin-to-skin contact, are not routinely used 
and have the potential to improve outcomes for preterm infants during the critical intrapartum and immediate newborn periods. 
The PTBi study was a CRT designed to improve the quality of care for mothers and preterm infants at the time of birth through a health facility  intervention (ClinicalTrials.gov: NCT03112018) \cite{ku_protocol}. The primary endpoint was intrapartum stillbirth and 28-day mortality among preterm infants delivered from October 2016 to May 2018 in Western Kenya and Eastern Uganda.

In more detail, 20 health facilities, including large hospitals and smaller health centers, were selected for participation in the study.
The facilities ranged in size, staff-to-patient ratio, and capacity to perform cesarean section (C-section), among others.
Prior to randomization, facilities were pair-matched 
{on country, delivery volume, staff-to-patient ratio, as well as rates of stillbirths, low-birth weight infants, and pre-discharge neonatal mortality} \cite{ptb_ku_walker}. Within matched pairs, they were than randomized to either the intervention or control arm \cite{ku_protocol}. 
Facilities in the control arm  received (1) strengthening of routine data collection and (2) introduction of the WHO Safe Childbirth Checklist \cite{WHO_checklist}. The facilities in the intervention arm received the components included in the control arm in addition to (1) PRONTO$^{TM}$  Simulation training \cite{pronto}, and (2) quality improvement collaboratives 
aimed to reinforce and optimize use of evidence-based practices. All study components consisted of known interventions and strategies aiming to improve quality of care, teamwork, communication, and data use \cite{ku_protocol}.

The results of the PTBi Study have been previously published \cite{ptb_ku_walker}.
Here, we focus on the real-world impact of highly variable cluster sizes when defining and estimating causal effect in CRTs. Specifically, during the study period, an unforeseen political strike led to lack of medical providers at certain facilities, thereby decreasing volume at some facilities while increasing volume at others. The number of preterm births for a given facility $N_j$ ranged between 40-366 in the intervention arm and between 29-447 in the control arm. Differences in cluster size were also pronounced within matched pairs and ranged between 9-211.

 To study the impact of high variability in cluster size, we return to the consequences of specifying the effect in terms of cluster-level outcomes (Eq.~\ref{Eq:ClusterRR}) versus individual-level outcomes (Eq.~\ref{Eq:IndvRR}). 
In PTBi, the individual-level outcome  $Y_{ij}$ was an indicator of preterm infant mortality by 28-day follow-up. Infants dying before discharge (stillbirth and predischarge mortality) were also included in the study; for these infants, $Y_{ij}=1$. 
For facility $j=\{1,\ldots, 20\}$, the cluster-level endpoint $Y^c_{j}$ was the  incidence of fresh stillbirth and 28-day all-cause mortality among preterm births and calculated as the empirical mean of the individual-level outcomes: $Y^c_j = 1/N_j \sum_{i=1}^{N_j} Y_{ij}$.

For both the cluster-level and individual-level effects, we compared estimates and inference from the TMLEs using cluster-level or individual-level data.  
The cluster-level TMLE used Adaptive Prespecification to select between no adjustment and adjustment for the proportion of mothers receiving a C-section, while the Hierarchical TMLE used Adaptive Prespecification to select between no adjustment and adjustment for an individual-level indicator of receiving a C-section.  
For comparison, we also implemented each TMLE without adjustment and each approach breaking or preserving the matched pairs used for randomization. (See Balzer et al. \cite{balzer_matching} for details on how matched analyses impact estimation and inference with TMLE.)  {Since C-section status was missing for 47 participants, all analyses restricted to the 2891 mother-infant dyads with complete data to improve comparability of the methods in this demonstration paper.}

\subsection{PTBi Results}

\begin{figure}[!h]
    \centering
    \includegraphics[width=0.75\textwidth]{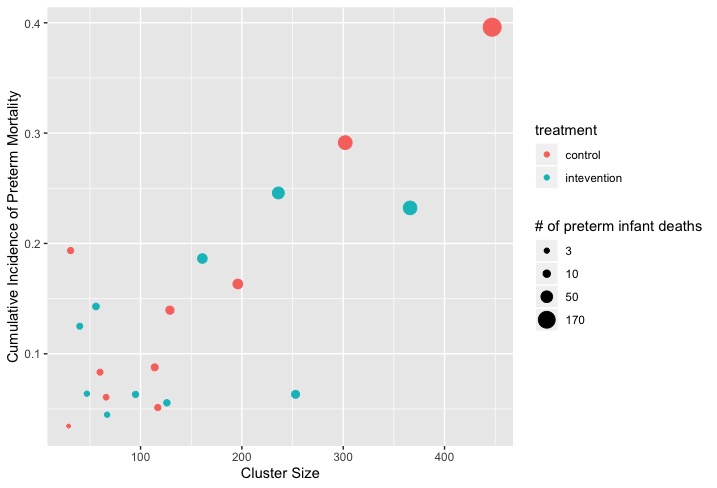}
    \caption{\textbf{Scatter plot of cluster size $N_j$ by the cumulative incidence of preterm mortality and trial arm in PTBi.}}
    \label{Fig:Scatterplot}
\end{figure}

As shown in \Cref{tab:adjusted_real_data_results_LB}, estimates and inferences for the effect of the PTBi intervention varied substantially by the target of inference.
At the cluster-level, the average incidence of 28-day mortality among preterm infants was 12\% among facilities randomized to the intervention and 15\% among facilities randomized to the control. Therefore, the unadjusted estimator of the cluster-level relative effect (Eq.~\ref{Eq:ClusterRR}) was 0.81
, corresponding to 19\% reduction in the  incidence of mortality at the facility-level. 
The results for the individual-level effect were markedly different, reflecting how outcome risk varied by cluster size. As shown in Figure~\ref{Fig:Scatterplot}, larger hospitals tended to have poorer outcomes (Pearson's correlation $r=0.77$), especially in the control arm ($r=0.88$). At an individual-level, the overall proportion of preterm infants who died within 28-days was 15\% in the intervention arm and 23\% in the control arm. Therefore, the unadjusted estimator of the individual-level relative effect (Eq.~\ref{Eq:IndvRR}) was 0.66
, corresponding to a 34\% reduction in the mortality risk.
As predicted by theory \cite{moultonhayes, imai2009, balzer_matching}, for a given statistical estimand, the analysis preserving the matched pairs  used for randomization was notably more precise than the analysis breaking the matches. Specifically, when keeping versus breaking the matches, the unadjusted estimator was 5-times more efficient for the cluster-level effect and 3-times more efficient  for the individual-level effect. Throughout, efficiency is defined as the variance of the unadjusted effect estimator when breaking the matches divided by the variance of another approach \cite{Balzer2021twostage}.

 Estimates from the TMLEs also indicated that the PTBi intervention reduced preterm mortality; again, the results varied by the target of inference and the estimation approach. For the cluster-level effect (Table~\ref{tab:adjusted_real_data_results_LB}-Left), adaptive adjustment for C-section doubled the precision of the analyses when breaking the matches. However, when preserving the matches, both the cluster-level TMLE and the Hierarchical TMLE defaulted to the unadjusted estimator, reflecting the ability of the adaptive approach to adjust only when it improves efficiency. Here, further adjustment for C-section status did not improve efficiency after matching on key outcome predictors (e.g., region and outcome rates prior to randomization).  Similar results are seen for the intervention effect at the individual-level effect (Table~\ref{tab:adjusted_real_data_results_LB}-Right). Adjustment for C-section improved the precision of analyses breaking the matches, but not analyses preserving the matches used for randomization.   It is worth noting that adjusted estimates tended to be of smaller magnitude than the unadjusted ones, especially for the cluster-level TMLE. This is not always the case \cite{Balzer2021twostage}, and we return to this in detail below.
 


\section{Discussion}

Cluster randomized trials (CRTs) are commonly used to evaluate the effects of interventions, which are randomized to groups of individuals, such as communities, clinics, or schools. 
In all trials, it is essential to clearly define the target causal effect and ensure the estimation approach follows from that effect \cite{ICHE9}. However, in CRTs, the hierarchical data structure (e.g., participants nested within clinics) complicates effect specification, especially when cluster sizes vary and interact with the intervention effect.
Additionally, in both resource-rich and resource-limited settings, high costs and implementation barriers often limit the number of clusters that can be enrolled and randomized. Therefore, it is important to understand the potential advantages and pitfalls of common analytic approaches in CRTs, overall and when there are few independent units of varying size.

In this manuscript, we aimed to provide a comprehensive framework for defining causal effects in CRTs and  an in-depth comparison of methods to estimate those effects. First, we used a structural causal model to describe the data generating process of a CRT and intervened upon it to generate counterfactual outcomes. Using these counterfactuals, we explored a variety of causal parameters that could be of interest, and we stressed the importance of \emph{a priori}-specifying the primary effect measure. We then delved into common CRT methods, and
considered each theoretically and with finite sample simulations. Finally, we demonstrated the impact of key analytic decisions using real data from the PTBi study, a clinic-based CRT designed to reduce mortality among preterm infants in East Africa. 

Our theoretical comparison of estimators revealed that common methods often target different causal effects. Results from our  simulation studies 
demonstrated how  targeted maximum likelihood estimation (TMLE) could improve statistical power by adaptively adjusting for covariates, while maintaining nominal confidence interval coverage. In the first simulation study, approaches based on generalized estimating equations (GEE and Aug-GEE) failed to preserve Type-I error control; {\color{black} this is consistent with the literature warning against their application with fewer than 30 clusters } \autocite{moultonhayes,murray}. However, it is worth noting that some studies may value minimizing Type-II error over Type-I error, and this should be carefully considered when developing the statistical analysis plan. 
Results from our second simulation study demonstrated the impact of informative cluster size, which occurs when the cluster size interacts with the intervention effect. In this setting, there can be a sharp divergence in the interpretation and magnitude of effects defined at the cluster-level or individual-level. This divergence was observed in the real data application, where  larger facilities in the control arm had poorer outcomes (Figure~\ref{Fig:Scatterplot}).

There are several limitations to the applied analysis, which merit additional consideration. 
{\color{black} Primarily, we only addressed two methodological challenges in  PTBi: defining and estimating causal effects in CRTs with few clusters of varying sizes. We did not address how to select candidate adjustment variables in practice \cite{adaptive_prespec,Benkeser2021}. 
Instead, due to limited data availability, our investigation explored adjustment for a single covariate: having a C-section at the individual-level or the proportion of women receiving a C-section at the cluster-level. 
This is a complex covariate in PTBi for the following reasons \cite{ptb_ku_walker}. First, not all health facilities had C-section capabilities at baseline. Further, the capacity to perform C-section changed for some facilities during the trial. Second, women with more complicated pregnancies were referred to facilities with C-section capacity, regardless of the randomization arm. However, during the trial, a political strike occurred, substantially impacting where women delivered --- again regardless of randomization arm. These real-world complexities reflect how CRTs are pragmatic and often need to respond to exogenous factors influencing both study interventions and outcomes. These secular events may entail modifications to the study design, follow-up period, primary endpoint, and statistical analysis plan; see, for example, Kakande et al. \cite{Kakande2022SIPT}.

As previously discussed, adjustment for C-section in PTBi attenuated estimates of the intervention effect, especially at the cluster-level. 
It is possible that the intervention impacted who and where women received C-sections --- thus violating the exclusion restriction in the causal model (Eq.~\ref{Eq:GeneralSCM}). While formal significance testing indicated no such effect, we cannot rule out that C-section might actually be a mediator. This reflects a  wider challenge, common in CRTs: how to define ``baseline'' when a longitudinal cohort of participants is not formed immediately after cluster-level randomization. Instead, in CRTs, participants may enroll or have their outcomes measured at various times after cluster-level randomization. Specifically in PTBi, mothers enrolled at the time of delivery, and infant outcomes were then measured from delivery until 28-days post-delivery. 
Addressing challenges with potential mediators, time-varying covariates, and missingness is beyond the scope of this review paper and can be handled using more complex methods \cite{diaz_missing, fiero,Balzer2020Supp, Balzer2021twostage, Leyrat2013, Leyrat2014, Li2022, NugentTB2022}.}

%

There are also limitations in our methods comparison. First, while we provided a comprehensive framework for defining causal effects, we did not provide explicit recommendations on whether to target the cluster-level or individual-level effect, on the additive or relative scale, and for the study units or a wider population. Instead, specification of the research query and the causal effect of interest must be made as a study team and will vary across real data applications. As we have demonstrated, TMLE provides flexibility to estimate a wide variety of user-specified effects in CRTs. In simulations, there was little practical difference in the performance of the cluster-level TMLE, using cluster-level data, versus  Hierarchical TMLE, using individual-level data, for the same statistical estimand. Both incorporated Adaptive Prespecification to select the optimal adjustment approach, given the data structure and target of inference. 
However, in the PTBi study, there were substantive differences in the point estimates and inference, when using the cluster-level TMLE versus Hierarchical TMLE. {\color{black}This may reflect that while both  cluster-level and Hierarchical TMLEs can target the same parameter,  the relationship between the cluster-level covariates and outcome may be distinct from the relationship between the individual-level covariates and outcome. Specifically, the impact of the proportion of women receiving a C-section on the cluster-level incidence of mortality is likely to be different than the impact of an individual woman's having a C-section on the risk of mortality for her child.}
Future work will extend Adaptive Prespecification to select between cluster-level and Hierarchical TMLEs, whichever maximizes empirical efficiency for the effect of interest. For now, the choice between cluster-level TMLE and Hierarchical TMLE (for the same parameter) may be driven by data availability and ease of implementation. 
Practically, it is most straightforward to estimate cluster-level effects with a cluster-level TMLE and individual-level effects with Hierarchical TMLE. However, in certain applications, data may only be available at the cluster-level, necessitating the use of cluster-level TMLE with weights to estimate individual-level effects.   
Altogether, to help inform the optimal analysis for a given application, we recommend conducting a simulation study reflecting the nuances of the real data problem.

In summary, CRTs are a popular approach for researchers seeking to estimate effects when interventions are scaled up to the population-level. 
Many of the challenges faced by the PTBi study are common  and highlight the importance of carefully selecting and prespecifying the primary causal effect, which should be driven by the trial's objectives \cite{ICHE9}.
For statistical estimation and inference, we recommend using TMLE given its flexibility to estimate a variety of effects and data-adaptively adjust for both cluster-level and individual-level covariates,  while preserving Type-I error control. 
We hope our presentation and comparison of various TMLEs has helped clarify its use and benefits in a CRT setting. Additionally, we have provided full \texttt{R} code in the Supplementary Materials and note that a \texttt{R} package for more general use of TMLE in randomized trials is under way \cite{tmle4RCTs}.

\section{Acknowledgments}
Thank you to the women and infants who
participated in the PTBi study. Thank you to the community health volunteers and the providers at each study facility in Kenya and Uganda. Thanks to the PTBi research teams at UCSF, Makerere University, and Kenya Medical Research Institute. Many thanks to Gertrude Namazzi, Kevin Achola, Christopher Otare, Paul Mubiri, Rikita Merai, Nancy Sloan, Lara Miller, Wenjing Zheng, and Caleb Miles. Thanks to the  Bill and Melinda Gates Foundation 
and National Institutes of Health for their generous support of this work.
 Lastly, we are very grateful for the mentors at UC Berkeley School of Public Health.
\section{Funding}
This work was supported by
the National Institutes of Health 
(U01AI150510, 
UM1AI068636, 
R01AI125000, 
R01AI074345, 
2R01AI074345-10A1). 
This work was also supported by the Bill \& Melinda Gates Foundation (OPP1107312). Under the grant conditions of the Foundation, a Creative Commons Attribution 4.0 Generic License has already been assigned to the Author Accepted Manuscript version that might arise from this submission. The funders did not play any direct role in study design, data collection, analysis, or writing of the manuscript.

\printbibliography

\newpage

\begin{table}[!htb]
    \caption{\textbf{ 
Performance of common CRT estimators when there is an effect and under the null across 500 iterations of Simulation I. }}
\label{tab:sim_I_res}
\centering
 \begin{tabular}{l cccccc | cccccc}\toprule
& \multicolumn{6}{c}{When there is an effect} & \multicolumn{6}{c}{Under the null} \\
 & pt  & bias & $\sigma$ & $\bar{\hat{\sigma}}$ & covg &  power & pt  & bias & $\sigma$ & $\bar{\hat{\sigma}}$ & covg &  Type-I \\
  \hline
Unadj & 0.84 & 0.01 & 0.17 & 0.17 & 0.97 & 0.18 & 1.01 & 0.01 & 0.16 & 0.16 & 0.96 & 0.04 \\ 
  C-TMLE & 0.83 & 0.00 & 0.04 & 0.05 & 0.98 & 0.99 & 1.00 & 0.00 & 0.03 & 0.03 & 0.98 & 0.02 \\ 
  H-TMLE & 0.83 & 0.00 & 0.04 & 0.05 & 0.98 & 0.99 & 1.00 & 0.00 & 0.03 & 0.04 & 0.98 & 0.02 \\ 
  t-test & 0.83 & 0.02 & 0.22 & 0.22 & 0.96 & 0.14 & 1.02 & 0.02 & 0.21 & 0.21 & 0.96 & 0.04 \\ 
  CARE & 0.83 & 0.02 & 0.05 & 0.05 & 0.86 & 0.93 & 1.00 & 0.00 & 0.03 & 0.03 & 0.96 & 0.04 \\ 
  GEE & 0.82 & -0.01 & 0.08 & 0.07 & 0.89 & 0.75 & 1.00 & 0.00 & 0.08 & 0.07 & 0.92 & 0.08 \\ 
  A-GEE & 0.83 & -0.00 & 0.09 & 0.06 & 0.84 & 0.77 & 1.00 & 0.00 & 0.09 & 0.06 & 0.84 & 0.16 \\ 
   \hline
\end{tabular}
\begin{tablenotes}
\footnotesize
\item ``Unadj'' refers to the unadjusted estimator, implemented as a cluster-level TMLE for the effect of interest and without covariate adjustment. ``C-TMLE'' and ``H-TMLE'' refer to the cluster-level TMLE and to Hierarchical TMLE, respectively; both were implemented with Adaptive Prespecification. ``A-GEE'' refers to Augmented-GEE.
\item ``pt" is the average point estimate; ``bias'' is the average difference between the point estimate and the target effect; $\sigma$ is the standard deviation of the point estimates on the log-scale; $\bar{\hat{\sigma}}$ is the average standard error estimate on the log-scale;  ``covg'' is the proportion of times the 95\% confidence interval contained the true effect, ``power'' is the proportion of times the false null hypothesis was rejected, and ``Type-I'' is the proportion of times the true null hypothesis was rejected.
\end{tablenotes}
\rule{8cm}{.5pt}
\end{table}

\newpage

\begin{table}[!htb]
    \caption{\textbf{ 
Performance of TMLEs for the cluster-level and individual-level relative effects across 500 iterations in Simulation II. }}
 \begin{tabular}{l cccccc | cccccc}\toprule
& \multicolumn{6}{c}{Cluster-level  effect: $\Psi^c_0(1)/\Psi^c_0(0)$ = 0.78} & \multicolumn{6}{c}{Individual-level  effect: $\Psi_0(1)/\Psi_0(0)$ = 0.69} \\
 & pt  & bias & $\sigma$ & $\bar{\hat{\sigma}}$ & covg &  power & pt  & bias & $\sigma$ & $\bar{\hat{\sigma}}$ & covg &  power\\
  \hline
 C-TMLE & 0.78 & 0.00 & 0.13 & 0.14 & 0.97 & 0.40 & 0.71 & 0.02 & 0.15 & 0.15 & 0.95 & 0.59 \\ 
  C-TMLE-AP & 0.79 & 0.01 & 0.12 & 0.12 & 0.96 & 0.44 & 0.71 & 0.02 & 0.14 & 0.14 & 0.94 & 0.68 \\ 
  H-TMLE & 0.78 & 0.00 & 0.12 & 0.14 & 0.98 & 0.40 & 0.71 & 0.02 & 0.15 & 0.16 & 0.96 & 0.60 \\ 
  H-TMLE-AP & 0.79 & 0.01 & 0.11 & 0.12 & 0.97 & 0.43 & 0.72 & 0.03 & 0.13 & 0.14 & 0.95 & 0.65 \\   \hline
\end{tabular}
\label{tab:ics_results}
\begin{tablenotes}
\footnotesize
\item 
``C-TMLE'' and ``H-TMLE'' refer to the cluster-level TMLE and to Hierarchical TMLE, respectively. Both were implemented with fixed  or adaptive adjustment via Adaptive Prespecification (``-AP'').
\item ``pt" is the average point estimate; ``bias'' is the average difference between the point estimate and the target effect; $\sigma$ is the standard deviation of the point estimates on the log-scale; $\bar{\hat{\sigma}}$ is the average standard error estimate on the log-scale;  ``covg'' is the proportion of times the 95\% confidence interval contained the true effect, and ``power'' is the proportion of times the false null hypothesis was rejected.
\end{tablenotes}
\rule{8cm}{1pt}
\end{table}

\newpage

\begin{table}[!htb]
\caption{\textbf{Estimating cluster-level and individual-level relative effects in the PTBi study.}}
    \label{tab:adjusted_real_data_results_LB}
        \centering
 \begin{tabular}[h]{l ccccc | ccccc}
\hline
 & \multicolumn{5}{c}{\textbf{For the cluster-level effect}} & 
        \multicolumn{5}{c}{\textbf{For the individual-level effect}} \\
& $\hat{\Psi}^c(1)$ & $\hat{\Psi}^c(0)$  & Ratio  (95\% CI)  & Eff. & Adj.
        & $\hat{\Psi}(1)$ & $\hat{\Psi}(0)$& Ratio  (95\% CI)  & Eff. & Adj. \\ 
\hline
 & \multicolumn{10}{c}{Breaking the matches} \\
Unadj. & 12\% & 15\% & 0.81 (0.43-1.55) & 1 & --- & 15\% & 23\% & 0.66 (0.33-1.31) & 1 &  --- \\ 
  C-TMLE & 13\% & 14\% & 0.88 (0.59-1.34) & 2 & C-sect & 17\% & 21\% & 0.84 (0.53-1.34) & 2 & C-sect \\ 
  H-TMLE & 12\% & 15\% & 0.84 (0.50-1.41) & 2 & C-sect & 16\% & 22\% & 0.70 (0.38-1.27) & 1 & C-sect \\ \hline
 & \multicolumn{10}{c}{Preserving the matches} \\
Unadj. & 12\% & 15\% & 0.81 (0.59-1.11) & 5 & --- & 15\% & 23\% & 0.66 (0.44-0.99) & 3 & ---\\ 
  C-TMLE & 12\% & 15\% & 0.81 (0.59-1.11) & 5 & $\emptyset$ & 15\% & 23\% & 0.65 (0.43-0.98) & 3 & $\emptyset$ \\ 
  H-TMLE & 12\% & 15\% & 0.82 (0.60-1.11) & 5 & $\emptyset$ & 16\% & 22\% & 0.70 (0.46-1.06) & 3 & C-sect \\ 
\hline
\end{tabular}
\begin{tablenotes}
\footnotesize
\item  ``Unadj.'' refers to the unadjusted effect estimator. ``C-TMLE'' and ``H-TMLE'' refer to the cluster-level TMLE and to Hierarchical TMLE, respectively; both were implemented with Adaptive Prespecification to adjust for C-section (``C-sect'') or nothing ($\emptyset$). ``Eff'' refers to the relative efficiency: the variance estimate for the unadjusted effect estimator breaking the matches used for randomization,
divided by the variance estimate of another approach (e.g., Hierarchical-TMLE with Adaptive Prespecification, keeping
the matches used for randomization).
\end{tablenotes}
\rule{8cm}{1pt}
\end{table}

\newpage

\newpage
\section*{Supplementary Materials}
\begin{center}
    \large{Defining and Estimating Effects in Cluster Randomized Trials:\\
A Methods Comparison}
\end{center}

\subfile{sections/supplement}

\end{document}

%% file: sections/supplement.tex
\maketitle

\section*{Appendix A: General Definition of Causal Effects in CRTs}

{\color{black}Recall $i=\{1,\ldots, N_j\}$ indexes individuals in cluster $j=\{1,\ldots,J\}$.}
Consider the cluster-level counterfactual outcome defined as the weighted sum of the individual-level counterfactual outcomes: 
$Y^c_j(a) = \sum_i \alpha_{j}Y_{ij}(a)$ for some cluster-specific weight
$\alpha_{j}$.  
We can further generalize our definition of treatment-specific mean as
\begin{align}
\Phi^{\star,J}(a) &=\frac{1}{J} \sum_{j=1}^J  \gamma_j Y_{j}^c(a)
\label{Eq:GammaRaysHulkAngry}
\end{align}
for user-specified weights 
 such that $\sum_j \gamma_j= J$. 
For example, setting $\gamma_j=1$ recovers the cluster-level parameter $\Phi^{c,J}(a)$, as in Eq.~\ref{Eq:Meta} {\color{black}in the Main Manuscript}. 
Alternatively, setting $\gamma_j=J/N_T \times 1/ \alpha_j$ recovers the individual-level parameter $\Phi^{J}(a)$, as in Eq.~\ref{Eq:IndvSample} {\color{black}in the Main Manuscript}.
To illustrate, we again focus on the setting where $\alpha_{j}=1/N_j$ is the inverse cluster size. Then we have 
\begin{align}
\frac{1}{J} \sum_{j=1}^J  \frac{J}{N_T} \frac{1}{\alpha_{j}} Y_{j}^c(a) = 
\frac{1}{J} \sum_{j=1}^J  \frac{J}{N_T} N_j \sum_{i=1}^{N_j} \frac{1}{N_j} Y_{ij}(a)
= \frac{1}{N_T} \sum_{j=1}^J  \sum_{i=1}^{N_j} Y_{ij}(a)
\label{Eq:BrainBreaker}
\end{align}
Altogether Eq. \ref{Eq:GammaRaysHulkAngry} allows us to consider a wide range of effects defined at the cluster-level or individual-level, regardless of whether the data are collected at the cluster-level or individual-level.

\section*{Appendix B: Hybrid TMLE}
\label{App:HybridTMLE}

Recall that the first step of the cluster-level TMLE (\Cref{section:cluster-level-estimators}) is to obtain an initial estimator of the conditional expectation of the cluster-level outcome $\mu^c(A,E,W^c)$. Instead of only considering cluster-level approaches, we can expand our candidate estimators by including aggregates of individual approaches \cite{hierarchical}. 
Consider, for example, the following specification of the expected individual-level outcome $\mathbb{E}(Y | A, E,W)$: \[
\mu(A,E,W) =  logit^{-1} [\beta_0 + \beta_A A + \beta_E E + \beta_W W ]
\]
We could estimate these coefficients by pooling participant data across clusters and running an individual-level logistic regression. Afterwards, for each cluster $j$, we would obtain and summarize the individual-level predicted outcomes to generate a candidate estimator of the expected cluster-level outcome: \[
\hat{\mu}^c(A_j,E_j,W^c_j) 
= \sum_{i=1}^{N_j} \alpha_{ij}\times  \hat{\mu}(A_j, E_j,W_{ij}) 
\]
with the selected $\alpha_{ij}$ corresponding to the relevant cluster-level summary of the individual-level outcomes (i.e., Eqs.~\ref{Eq:ClusterCF} and~\ref{eq:cluster_y} {\color{black}in the Main Manuscript}); throughout, we have been focusing on $\alpha_{ij}=1/N_j$ for $i=\{1,\ldots, N_j\}$.
Then estimation and inference would proceed as described in \Cref{section:cluster-level-estimators} for the cluster-level TMLE.  
Thus, this approach naturally targets the cluster-level treatment-specific mean $\Psi^c_0(a)=\mathbb{E}_0[\mathbb{E}_0(Y^c|a,E,W^c)]$, 
and therefore  cluster-level effects, as in Eq.~\ref{Eq:ClusterRR} {\color{black}in the Main Manuscript}. However, as previously discussed, we can modify the weights to, instead, target an individual-level effect.
More importantly, we can now use Adaptive Prespecification to choose between candidate estimators of $\mu^c$ based on cluster-level approaches or aggregates of individual-level approaches. In the latter, the initial estimator of $\mu^c$ is based on individual-level data that is pooled across clusters; therefore, we could consider methods that are more data-adaptive than GLMs. Specifically, 
we could use Super Learner, an ensemble machine learning algorithm \cite{superlearner_2007}.
Additionally, since the cluster-level TMLE (Section~\ref{section:cluster-tmle-adapt}), the Hierarchical TMLE (Section~\ref{section:hier_tmle}), and the Hybrid TMLE (described here) can target the same effects, we can use Adaptive Prespecification to select the TMLE resulting in the greatest precision for the effect of interest.

\section*{Appendix C: Additional Results for Simulation I}

Recall in Simulation I, the cluster sizes varied, but the cluster-level relative effect was equivalent to the individual-level relative effect: 0.83. In other words, there was no informative cluster size. While the true values of target effects were identical, their interpretation (Sec.~\ref{Section:effects}) and estimator implementation  depends on the target. Therefore, we expanded Simulation I to study the finite sample performance of the unadjusted estimator, the cluster-level TMLE, and Hierarchical TMLE when the target of inference was the cluster-level and individual-level relative effect. Additionally, we examined performance with fewer clusters ($J=10$), smaller clusters ($N_j \sim Norm(20,10)$ subject to a minimum of 10 participants/cluster), and under the null, where we generated the outcome when setting the intervention terms to zero:
  $Y_{ij} \sim \mathbbm{1}\big[ U_{Y_{ij}} < logit^{-1}(-0.75 + 0 \times A_j + 0.8W1_{ij} + 0.4W2_{ij} - 0.3E1_j + 0\times A_j W2_{ij}) \big]$.
 
Across all settings and as expected, the attained power of the TMLEs was substantially higher than the unadjusted effect estimator (\Cref{tab:sim_1_appendix_EFFECT} of the Supplementary Materials). Also as expected, a smaller number of randomized units $J$ and smaller cluster sizes (i.e., average $N_j$) resulted in reduced power. Throughout, the TMLEs maintained good-to-conservative confidence interval coverage. Under the null, the TMLEs also had good-to-conservative Type-I error control in all settings (\Cref{tab:sim_1_appendix_NULL} of the Supplementary Materials).
Altogether these additional simulation results illustrate how TMLE can flexibly estimate a variety of causal effects and adaptively adjust for substantial precision gains, while preserving confidence interval coverage and Type-I error control.

\begin{table}[!h]
    \caption{\textbf{ 
Performance of TMLEs for the cluster-level and individual-level relative effects across 500 iterations of Simulation I when there is an effect.}}
 \begin{tabular}{lrrrrrr | rrrrrr}\toprule
  & \multicolumn{6}{c}{Cluster-level  effect: $\Psi^c_0(1)/\Psi^c_0(0)$ = 0.83} & \multicolumn{6}{c}{Individual-level  effect: $\Psi_0(1)/\Psi_0(0)$ = 0.83} \\
 & pt  & bias & $\sigma$ & $\bar{\hat{\sigma}}$ & covg &  power & pt  & bias & $\sigma$ & $\bar{\hat{\sigma}}$ & covg &  power \\   
 & \multicolumn{12}{c}{$J=20$ clusters of average size $N_j$=150} \\
Unadj & 0.84 & 0.01 & 0.17 & 0.17 & 0.97 & 0.18 & 0.85 & 0.02 & 0.20 & 0.18 & 0.95 & 0.17 \\ 
  C-TMLE & 0.83 & 0.00 & 0.04 & 0.05 & 0.98 & 0.99 & 0.83 & -0.00 & 0.04 & 0.05 & 0.97 & 0.98 \\ 
  H-TMLE & 0.83 & 0.00 & 0.04 & 0.05 & 0.98 & 0.99 & 0.83 & -0.00 & 0.04 & 0.05 & 0.97 & 0.97 \\ 
   & \multicolumn{12}{c}{$J=10$ clusters of average size $N_j$=150} \\
  Unadj & 0.85 & 0.03 & 0.26 & 0.24 & 0.96 & 0.10 & 0.87 & 0.04 & 0.29 & 0.24 & 0.93 & 0.12 \\ 
  C-TMLE & 0.83 & -0.00 & 0.06 & 0.06 & 0.97 & 0.79 & 0.83 & -0.00 & 0.09 & 0.06 & 0.95 & 0.78 \\ 
  H-TMLE & 0.83 & -0.00 & 0.06 & 0.07 & 0.97 & 0.75 & 0.83 & -0.00 & 0.07 & 0.07 & 0.95 & 0.74 \\ 
   & \multicolumn{12}{c}{$J=20$ clusters of average size $N_j$=20} \\
Unadj & 0.85 & 0.01 & 0.19 & 0.19 & 0.95 & 0.14 & 0.85 & 0.01 & 0.19 & 0.19 & 0.95 & 0.13 \\ 
  C-TMLE & 0.83 & -0.00 & 0.07 & 0.07 & 0.96 & 0.69 & 0.83 & -0.00 & 0.07 & 0.07 & 0.96 & 0.73 \\ 
  H-TMLE & 0.83 & -0.00 & 0.07 & 0.07 & 0.97 & 0.68 & 0.83 & -0.00 & 0.07 & 0.07 & 0.96 & 0.71 \\
 & \multicolumn{12}{c}{$J=10$ clusters of average size $N_j$=20} \\
Unadj & 0.86 & 0.03 & 0.27 & 0.26 & 0.96 & 0.08 & 0.87 & 0.04 & 0.28 & 0.26 & 0.96 & 0.09 \\ 
  C-TMLE & 0.83 & -0.01 & 0.11 & 0.10 & 0.94 & 0.39 & 0.83 & -0.00 & 0.10 & 0.09 & 0.94 & 0.44 \\ 
  H-TMLE & 0.83 & -0.01 & 0.11 & 0.10 & 0.95 & 0.39 & 0.83 & -0.00 & 0.10 & 0.09 & 0.94 & 0.41 \\ 
  \bottomrule
\end{tabular}
\label{tab:sim_1_appendix_EFFECT}
\begin{tablenotes}
\footnotesize
\item ``Unadj'' refers to the unadjusted estimator, implemented as a cluster-level TMLE for the effect of interest and without covariate adjustment. 
``C-TMLE'' and ``H-TMLE'' refer to the cluster-level TMLE and to Hierarchical TMLE, respectively; both were implemented with Adaptive Prespecification to select from $\{\emptyset, W1^c, W2^c\}$ and $\{\emptyset, W1, W2\}$, respectively.
\item ``pt" is the average point estimate; ``bias'' is the average difference between the point estimate and the target effect; $\sigma$ is the standard deviation of the point estimates on the log-scale; $\bar{\hat{\sigma}}$ is the average standard error estimate on the log-scale;  ``covg'' is the proportion of times the 95\% confidence interval contained the true effect, ``power'' is the proportion of times the false null hypothesis was rejected.
\end{tablenotes}
\rule{8cm}{1pt}
\end{table}

\begin{table}[!h]
    \caption{\textbf{ 
Performance of TMLEs for the cluster-level and individual-level relative effects across 500 iterations of Simulation I under the null}}
 \begin{tabular}{lrrrrrr | rrrrrr}\toprule
 & \multicolumn{6}{c}{Cluster-level  effect: $\Psi^c_0(1)/\Psi^c_0(0)$ =1.0} & \multicolumn{6}{c}{Individual-level  effect: $\Psi_0(1)/\Psi_0(0)$ = 1.0} \\
\textbf{Null} & pt  & bias & $\sigma$ & $\bar{\hat{\sigma}}$ & covg &  Type-I & pt  & bias & $\sigma$ & $\bar{\hat{\sigma}}$ & covg &  Type-I \\
 & \multicolumn{12}{c}{$J=20$ clusters of average size $N_j$=150} \\
 Unadj & 1.01 & 0.01 & 0.16 & 0.16 & 0.96 & 0.04 & 1.02 & 0.02 & 0.18 & 0.17 & 0.95 & 0.05 \\ 
  C-TMLE & 1.00 & 0.00 & 0.03 & 0.04 & 0.99 & 0.01 & 1.00 & -0.00 & 0.03 & 0.03 & 0.98 & 0.02 \\ 
  H-TMLE & 1.00 & 0.00 & 0.03 & 0.04 & 0.98 & 0.02 & 1.00 & -0.00 & 0.03 & 0.04 & 0.98 & 0.02 \\ 
   & \multicolumn{12}{c}{$J=10$ clusters of average size $N_j$=150} \\
Unadj & 1.03 & 0.03 & 0.24 & 0.22 & 0.97 & 0.03 & 1.05 & 0.05 & 0.27 & 0.23 & 0.93 & 0.07 \\ 
  C-TMLE & 1.00 & 0.00 & 0.05 & 0.05 & 0.98 & 0.02 & 1.00 & 0.00 & 0.05 & 0.04 & 0.97 & 0.03 \\ 
  H-TMLE & 1.00 & 0.00 & 0.05 & 0.05 & 0.98 & 0.02 & 1.00 & 0.00 & 0.05 & 0.05 & 0.97 & 0.03 \\ 
   & \multicolumn{12}{c}{$J=20$ clusters of average size $N_j$=20} \\
Unadj & 1.01 & 0.01 & 0.17 & 0.17 & 0.94 & 0.06 & 1.02 & 0.02 & 0.18 & 0.18 & 0.95 & 0.05 \\ 
  C-TMLE & 1.00 & -0.00 & 0.06 & 0.06 & 0.97 & 0.03 & 1.00 & -0.00 & 0.06 & 0.05 & 0.96 & 0.04 \\ 
  H-TMLE & 1.00 & -0.00 & 0.06 & 0.06 & 0.97 & 0.03 & 1.00 & -0.00 & 0.06 & 0.06 & 0.96 & 0.04 \\ 
 & \multicolumn{12}{c}{$J=10$ clusters of average size $N_j$=20} \\
Unadj & 1.04 & 0.04 & 0.25 & 0.24 & 0.97 & 0.03 & 1.05 & 0.05 & 0.26 & 0.24 & 0.97 & 0.03 \\ 
  C-TMLE & 1.00 & 0.00 & 0.09 & 0.08 & 0.94 & 0.06 & 1.00 & 0.00 & 0.08 & 0.07 & 0.94 & 0.06 \\ 
  H-TMLE & 1.00 & 0.00 & 0.09 & 0.08 & 0.95 & 0.05 & 1.00 & 0.00 & 0.08 & 0.07 & 0.94 & 0.06 \\
  \bottomrule
\end{tabular}
\label{tab:sim_1_appendix_NULL}
\begin{tablenotes}
\footnotesize
\item  ``Unadj'' refers to the unadjusted estimator, implemented as a cluster-level TMLE for the effect of interest and without covariate adjustment.  
``C-TMLE'' and ``H-TMLE'' refer to the cluster-level TMLE and to Hierarchical TMLE, respectively; both were implemented with Adaptive Prespecification to select from $\{\emptyset, W1^c, W2^c\}$ and $\{\emptyset, W1, W2\}$, respectively.
\item ``pt" is the average point estimate; ``bias'' is the average difference between the point estimate and the target effect; $\sigma$ is the standard deviation of the point estimates on the log-scale; $\bar{\hat{\sigma}}$ is the average standard error estimate on the log-scale;  ``covg'' is the proportion of times the 95\% confidence interval contained the true effect,  and ``Type-I'' is the proportion of times the true null hypothesis was rejected.
\end{tablenotes}
\rule{8cm}{1pt}
\end{table}


\section*{Appendix D: Additional Results from Simulation II}

 In Simulation II, we examined performance under the null, where we generated the outcome by setting the intervention terms to zero:   $Y_{ij} \sim \mathbbm{1}[U_{Y_{ij}} < logit ^ {-1} 
      (0.5 + W1_{ij}/6 + W2_{ij}/2 + W3_{ij}/4 + E1_j/5 + E2_j/5 - \tilde{N}_j/8 - 0 \times A\tilde{N}_j/5 )]$.
The results are given in  \Cref{tab:sim_2_appendix_breaking_matches} of the Supplementary Materials and demonstrate the TMLEs with and without Adaptive Prespecification tightly preserved Type-I error control for both the cluster-level and individual-level effect.

\begin{table}[!h]
    \caption{\textbf{ 
Performance of TMLEs for the cluster-level and individual-level relative effects across 500 iterations in Simulation II under the null}}
 \begin{tabular}{lrrrrrr | rrrrrr}\toprule
& \multicolumn{6}{c}{Cluster-level  effect: $\Psi^c_0(1)/\Psi^c_0(0)$ = 1} & \multicolumn{6}{c}{Individual-level  effect: $\Psi_0(1)/\Psi_0(0)$ = 1} \\
 & pt  & bias & $\sigma$ & $\bar{\hat{\sigma}}$ & covg &  Type-I & pt  & bias & $\sigma$ & $\bar{\hat{\sigma}}$ & covg &  Type-I\\
  \hline
C-TMLE & 1.01 & 0.01 & 0.09 & 0.10 & 0.98 & 0.02 & 1.01 & 0.01 & 0.11 & 0.12 & 0.96 & 0.04 \\ 
  C-TMLE-AP & 1.01 & 0.01 & 0.07 & 0.09 & 0.98 & 0.02 & 1.01 & 0.01 & 0.09 & 0.10 & 0.96 & 0.04 \\ 
  H-TMLE& 1.01 & 0.01 & 0.09 & 0.11 & 0.98 & 0.02 & 1.01 & 0.01 & 0.10 & 0.12 & 0.97 & 0.03 \\ 
  H-TMLE-AP & 1.01 & 0.01 & 0.07 & 0.09 & 0.99 & 0.01 & 1.01 & 0.01 & 0.08 & 0.10 & 0.98 & 0.02 \\ 
   \hline
\end{tabular}
\label{tab:sim_2_appendix_breaking_matches}
\begin{tablenotes}
\footnotesize
\item 
``C-TMLE'' and ``H-TMLE'' refer to the cluster-level TMLE and to Hierarchical TMLE, respectively. Both were implemented with fixed  or adaptive adjustment via Adaptive Prespecification (``-AP'').
\item ``pt" is the average point estimate; ``bias'' is the average difference between the point estimate and the target effect; $\sigma$ is the standard deviation of the point estimates on the log-scale; $\bar{\hat{\sigma}}$ is the average standard error estimate on the log-scale;  ``covg'' is the proportion of times the 95\% confidence interval contained the true effect, and ``Type-I'' is the proportion of times the true null hypothesis was rejected.
\end{tablenotes}
\rule{8cm}{1pt}
\end{table}

\section*{Appendix E: Computing code}
The simulation study and real data analysis were conducted in \texttt{R} (v4.2.1) and utilized the \texttt{nbpMatching}, \texttt{geesmv}, \texttt{CRTgeeDR}, and \texttt{ltmle}
packages, among others \cite{R, nbp_match, geesmv, CRTgeeDR, ltmlepackage}. Computing code to reproduce the simulation studies and to analyze the PTBi Study is available at
\url{https://github.com/LauraBalzer/Comparing_CRT_Methods}. As previously noted, an \texttt{R} package to estimate effects with TMLE in both individually randomized and cluster randomized trials is under construction: \url{https://github.com/LauraBalzer/tmle4rcts}.

\clearpage

\newpage

%% file: comp_methods.bib
@article{Balzer2020Supp,
	author = {L.B. Balzer and J. Ayieko and D. Kwarisiima and G. Chamie and E.D. Charlebois and J. Schwab and M.J. {van der Laan} and M.R. Kamya and D.V. Havlir and M.L. Petersen},
	journal = {Epidemiology},
	number = {5},
	pages = {620-627},
	title = {Far from {MCAR}: obtaining population-level estimates of {HIV} viral suppression},
	volume = {31},
	year = {2020}}

@article{Kakande2022SIPT,
	author = {E. Kakande and C. Christian and L.B. Balzer and A. Owaraganise and others},
	journal = {Lancet HIV},
	number = {9},
	pages = {E607-E616},
	title = {A mid-level health manager intervention to promote uptake of {I}soniazid {P}reventive {T}herapy in {U}ganda: a cluster randomized trial},
	volume = {9},
	year = {2022}}

@book{RothmanModern,
	address = {Phildelphia},
	author = {Rothman, K.J. and Greenland, S. and Lash, T.L.},
	edition = {3rd},
	publisher = {Lippincott Williams \& Wilkins},
	title = {Modern Epidemiology},
	year = {2008}}

@article{Balzer2022FancyAPS,
	author = {L.B. Balzer and E. Cai and L. {Godoy Garraza} and P. Amaranath},
	journal = {https://arxiv.org/abs/2210.17453},
	title = {Adaptive Selection of the Optimal Strategy to Improve Precision and Power in Randomized Trials},
	year = {2022}}

@article{Li2022,
	author = {Fan Li and Zizhong Tian and Jennifer Bobb and Georgia Papadogeorgou and Fan Li},
	journal = {Clinical Trials},
	number = {1},
	pages = {33-41},
	title = {Clarifying selection bias in cluster randomized trials},
	volume = {19},
	year = {2022}}

@article{Imbens2004,
	author = {G.W. Imbens},
	doi = {10.1162/003465304323023651},
	journal = {Review of Economics and Statistics},
	number = {1},
	pages = {4-29},
	title = {Nonparametric estimation of average treatment effects under exogeneity: a review},
	volume = {86},
	year = {2004}}

@article{Balzer2017,
	author = {L.B. Balzer},
	journal = {Epidemiology},
	number = {4},
	pages = {562--566},
	title = {{``All} generalizations are dangerous, even this one.'' - {Alexandre} {Dumas} {[Commentary]}},
	volume = {28},
	year = {2017}}

@article{Li2016,
	author = {F. Li and Y. Lokhnygina and D. Murray and P.J. Heagerty and E.R. DeLong},
	journal = {Statistics in medicine},
	number = {10},
	pages = {1565-1579},
	title = {An evaluation of constrained randomization for the design and analysis of group randomized trials},
	volume = {35},
	year = {2016}}

@article{Li2017,
	author = {F. Li and E.L. Turner and P.J. Heagerty and D.M. Murray and W.M. Vollmer and E. DeLong},
	journal = {Statistics in medicine},
	number = {24},
	pages = {3791-3806},
	title = {An evaluation of constrained randomization for the design and analysis of group randomized trials with binary outcomes},
	volume = {36},
	year = {2017}}

@article{Leyrat2014,
	author = {C. Leyrat and A. Caille and  A. Donner and B. Giraudeau},
	journal = {Statistics in medicine},
	number = {20},
	pages = {3556-3575},
	title = {Propensity score methods for estimating relative risks in cluster randomized trials with low incidence binary outcomes and selection bias},
	volume = {33},
	year = {2014}}

@article{Leyrat2013,
	author = {Leyrat, C. and Caille, A. and Donner, A. and Giraudeau, B.},
	journal = {Statistics in Medicine},
	number = {19},
	pages = {3357-3372},
	title = {Propensity scores used for analysis of cluster randomized trials with selection bias: a simulation study},
	volume = {32},
	year = {2013}}

@article{Ding2021,
	author = {F. Su and P. Ding},
	journal = {Journal of the Royal Statistical Society: Series B (Statistical Methodology)},
	number = {5},
	pages = {994-1015},
	title = {Model‐assisted analyses of cluster‐randomized experiments},
	volume = {83},
	year = {2021}}

@article{Cook2016,
	author = {A. J. Cook and E. Delong and D. Murray and W. Vollmer and P. J Heagerty},
	journal = {Clinical Trials},
	number = {5},
	pages = {504-512},
	title = {Statistical Lessons Learned for Designing Cluster Randomized Pragmatic Clinical Trials from the NIH Health Care Systems Collaboratory Biostatistics and Design Core},
	volume = {13},
	year = {2016}}

@article{NugentTB2022,
	author = {J.R. Nugent and C. Marquez and E. Charlebois and R. Abbott and L.B. Balzer and for the {SEARCH} Collaboration},
	journal = {https://arxiv.org/abs/2208.09508},
	title = {Blurring cluster randomized trials and observational studies: Two-Stage {TMLE} for sub-sampling, missingness, and few independent units},
	year = {2022}}

@manual{tmle4RCTs,
	author = {L.B. Balzer and J. Nugent},
	title = {tmle4CRTs: {TMLE} for randomized trials},
	year = {2022}}

@article{Benkeser2021,
	author = {D. Benkeser and I. D\'{i}az and A. Luedtke and J. Segal and D. Scharfstein and M. Rosenblum},
	journal = {Biometrics},
	number = {4},
	pages = {1467-1481},
	title = {Improving Precision and Power in Randomized Trials for {COVID-19} Treatments Using Covariate Adjustment, for Binary, Ordinal, and Time-to-Event Outcomes},
	volume = {77},
	year = {2021}}

@Manual{geesmv,
    title = {geesmv: Modified Variance Estimators for Generalized Estimating
Equations},
    author = {Ming Wang},
    year = {2015},
    note = {R package version 1.3},
    url = {https://CRAN.R-project.org/package=geesmv},
  }

@article{Fay2001,
	author = {M.P. Fay and B.I. Graubard },
	journal = {Biometrics},
	pages = {1198-1206},
	title = {Small-sample adjustments for {W}ald-type tests using sandwich estimators},
	volume = {57},
	year = {2001}}

@article{Havlir2019,
	author = {D.V. Havlir and L.B. Balzer and E. Charlebois and T.D. Clark and D. Kwarisiima and J. Ayieko and J Kabami and N. Sang and others},
	journal = {New England Journal of Medicine},
	pages = {219--229},
	title = {{HIV} testing and treatment with the use of a community health approach in rural {A}frica},
	volume = {381},
	year = {2019}}

@article{Wang2021weightGEE,
	author = {X. Wang and E.L. Turner and F. Li and R. Wang and J. Moyer and A.J. Cook and D.M. Murray and P.J. Heagerty},
		journal = {Contemporary Clinical Trials},
	pages = {p.106702},
	title = {Two weights make a wrong: cluster randomized trials with variable cluster sizes and heterogeneous treatment effects},
	volume = {114},
	year = {2022}}

@manual{R,
	Address = {Vienna, Austria},
	Author = {{R Core Team}},
	Organization = {R Foundation for Statistical Computing},
	Title = {R: A Language and Environment for Statistical Computing},
	Url = {http://www.R-project.org},
	Year = 2020}

@article{Kahan2021,
	Author = {B.C. Kahan and T.P. Morris and I.R. White and J. Carpenter and S. Cro},
	Journal = {Trials},
	Number = {686},
	Pages = {https://doi.org/10.1186/s13063-021-05644-4},
	Title = {Estimands in published protocols of randomised trials: urgent improvement needed},
	Volume = {22},
	Year = {2021}}

@techreport{ICHE9,
	Author = {{International Council for Harmonisation}},
	Institution = {European Medicines Agency},
	Title = {{ICH E9 (R1)} addendum on estimands and sensitivity analysis in clinical trials to the guideline on statistical principles for clinical trials},
	Type = {https://www.ema.europa.eu/en/ich-e9-statistical-principles-clinical-trials},
	Year = {2020}}

@manual{CRTgeeDR,
	Author = {M. Prague},
	Organization = {R package version 2.0},
	Title = {CRTgeeDR: Doubly Robust Inverse Probability Weighted Augmented GEE Estimator},
	Url = {https://CRAN.R-project.org/package=CRTgeeDR},
	Year = {2017}}

@article{WHO_checklist,
    Author = {Spector JM and Lashoher A and Agrawal P and Claire Lemer C and Dziekan G and Bahl R and Matthews M Mathai and Merialdi M and Berry W and Gawande AA}, 
    Title = {Designing the WHO Safe Childbirth Checklist program to improve quality of care at childbirth.},
    Journal = {International Journal of Gynecology \& Obstetrics},
    Year = {2013},
    Volume = {122},
    Number = {2},
    Pages = {164-168}
}

@article{ptb_ku_walker,
    Author = {Dilys Walker and Phelgona Otieno and Elizabeth Butrick and Gertride Namazzi and Achola K and Merai R and Otare C and Mubiri P and Ghosh R and Santos N and Miller L and Sloan NL and Waiswa P},
    Title = {Preterm Birth Initiative Kenya and Uganda Implementation Research Collaborative. Effect of a quality improvement package for intrapartum and immediate newborn care on fresh stillbirth and neonatal mortality among preterm and low-birthweight babies in Kenya and Uganda: a cluster-randomised facility-based trial.},
    Journal = {Lancet Glob Health.},
    Year = {2020},  
    Volume = {8},
    Number = {8},
    Pages={e1061-e1070}}

@article{nevail_ics2014,
	Author = {Jaakko Nevalainen and S. Datta and H. Oja},
	Journal = {Statistical Papers},
	Title = {Inference on the marginal distribution of clustered data with informative cluster size},
	Volume = {55},
	Number={1},
	Pages={71-92},
	Year = {2014}}

@article{BalzerMLcomm2021,
	Author = {L.B. Balzer and M.L. Petersen},
	Journal = {Am J Epidemiol},
	Title = {Machine Learning in Causal Inference: \emph{How do I love thee? Let me count the ways.}},
	Volume = {190},
	Number={8},
	Pages = {1483-1487},
	Year = {2021}}

@article{Petersen2014roadmap,
	Author = {M.L. Petersen and M.J. van der Laan},
	Journal = {Epidemiology},
	Number = {3},
	Pages = {418-426},
	Title = {Causal Models and Learning from Data: Integrating Causal Modeling and Statistical Estimation},
	Volume = {25},
	Year = {2014}}

@article{ltmlepackage,
	Author = {S.D. Lendle and J. Schwab and M.L. Petersen and M.J. {van der Laan}},
	Journal = {Journal of Statistical Software},
	Number = {1},
	Pages = {1--21},
	Title = {ltmle: An {R} Package Implementing Targeted Minimum Loss-based Estimation for Longitudinal Data},
	Volume = {81},
	Year = {2017}}

@article{Schnitzer2014,
	Author = {M.E. Schnitzer and M.J. van der Laan and E.E. Moodie and R.W. Platt},
	Journal = {Annals of Applied Statistics},
	Number = {2},
	Pages = {703--725},
	Title = {Effect of breastfeeding on gastrointestinal infection in infants: a targeted maximum likelihood approach for clustered longitudinal data},
	Volume = {8},
		Year = {2014}}

@article{Murray2020,
	Author = {D.M. Murray and M. Taljaard and E.L. Turner and S.M. George},
	Journal = {Annu Rev Public Health},
	Pages = {1-19},
	Title = {Essential Ingredients and Innovations in the Design and Analysis of Group-Randomized Trials},
	Volume = {41},
	Year = {2020}}

@article{Balzer2021twostage,
	Author = {L.B. Balzer and M. {van der Laan} and J. Ayieko$^*$ and M. Kamya and others},
	Journal = {Biostatistics},
	Title = {Two-Stage {TMLE} to Reduce Bias and Improve Efficiency in Cluster Randomized Trials},
	Volume = {kxab043},
	Year = {2021}}

@article{Moore2009,
	Author = {Moore, K.L. and van der Laan, M.J.},
	Doi = {10.1002/sim.3445},
	Journal = {Statistics in Medicine},
	Number = {1},
	Pages = {39--64},
	Title = {Covariate Adjustment in Randomized Trials with Binary Outcomes: {T}argeted Maximum Likelihood Estimation},
	Volume = {28},
	Year = {2009}}

@article{Seaman2014,
	Author = {S.R. Seaman and M. Pavlou and A.J. Copas},
	Journal = {Statistics in Medicine},
	Pages = {5371--5387},
	Title = {Review of methods for handling confounding by cluster and informative cluster size in clustered data},
	Volume = {33},
	Year = {2014}}

@article{aug-gee,
	Author = {Alisa J. Stephens and Eric J. Tchetgen Tchetgen and Victor de Gruttola},
	Journal = {Statistics in Medicine},
	Number = {10},
	Pages = {915-930},
	Title = {Augmented GEE for improving efficiency and validity of estimation in cluster randomized trials by leveraging cluster and individual-level covariates},
	Volume = {31},
	Year = {2012}}

@article{stephens_aug-gee14,
	Author = {Alisa J. Stephens and Eric J. Tchetgen Tchetgen and Victor de Gruttola},
	Journal = {The International Journal of Biostatistics},
	Volume = {10},
	Number = {1},
	Pages = {59-75},
	Title = {Locally efficient estimation of marginal treatment effects when outcomes are correlated: is the prize worth the chase?},
	Year = {2014}}

@article{gee_correction,
	Author = {Peng Li and David T. Redden},
	Journal = {Statistics in Medicine},
	Number = {2},
	Pages = {281-296},
	Title = {Small Sample Performance of Bias-corrected Sandwich Estimators for Cluster-Randomized Trials with Binary Outcomes},
	Volume = {34},
	Year = {2015}}

@article{liang-zeger,
	Author = {Kung-Yee Liang and Scott L. Zeger},
	Journal = {Biometrika},
	Number = {1},
	Pages = {13-22},
	Title = {Longitudinal Data Analysis Using Generalized Linear Model},
	Volume = {73},
	Year = {1986}}

@book{ware-laird,
	Author = {Garrett M. Fitzmaurice and Nan M. Laird and James H. Ware},
	Edition = {second},
	Publisher = {Wiley},
	Title = {Applied Longitudinal Analysis},
	Year = {2011}}

@article{hubbard_ahern,
	Author = {Alan E. Hubbard and Jennifer Ahern and Nancy L. Fleischer and Mark van der Laan and Sheri A. Lippman and Nicholas Jewell and Tim Pruckner and William A. Satariano},
	Journal = {Epidemiology},
	Pages = {467-474},
	Title = {To GEE or Not to GEE: Comparing Population Average and Mixed Models for Estimating the Associations Between Neighborhood Risk Factors and Health},
	Volume = {21},
	Number = {4},
	Year = {2010}}

@article{gail-care,
	Author = {Mitchell Gail and Steven Mark and Sylvan B. Green and David Pee},
	Journal = {Statistics in Medicine},
	Pages = {1069-1092},
	Title = {On design considerations and randomization-based inference for community intervention trials},
	Volume = {15},
	Year = {1996}}

@article{hierarchical,
	Author = {Laura Balzer and Wenjing Zheng and Mark J. van der Laan and Maya Petersen},
	Journal = {Statistical Methods in Medical Research},
	Title = {A new approach to hierarchical data analysis: Targeted maximum likelihood estimation for the causal effect of a cluster-level exposure},
	Year = {2019},
	Volume=28,
	Number=6,
	Pages={1761-1780}}

@article{adaptive_prespec,
	Author = {Laura Balzer and Mark J. van der Laan and Maya Petersen},
	Journal = {Statistics in Medicine},
	Title = {Adaptive pre-specification in randomized trials with and without pair-matching},
	Year = {2016},
	Volume=35, 
	Pages={4528-4545} }

@book{moultonhayes,
	Author = {Richard J. Hayes and Lawrence Hale Moulton},
	Edition = {2nd edition},
	Publisher = {Chapman \& Hall/CRC},
	Title = {Cluster randomised trials},
	Year = {2017}}

@book{rose_tl,
	Author = {Mark J. van der Laan and Sherri Rose},
	Edition = {1st edition},
	Publisher = {Springer Science+Business Media},
	Title = {Targeted learning: causal inference for observational and experimental data},
	Year = {2011}}

@article{icc,
	Author = {Christina Pagel and Audrey Prost and Sonia Lewycka and Sushmita Das and Tim Colbourn and Rajendra Mahapatra and Kishwar Azad and Anthony Costello and David Osrin},
	Journal = {Trials},
	Title = {Intracluster correlation coefficients and coefficients of variation for perinatal outcomes from five cluster-randomised controlled trials in low and middle-income countries: results and methodological implications},
	Year = {2011},
	Volume = {12},
	Number = {151}}

@article{murray,
	Author = {David M. Murray and Sherri L. Pals and Stephanie M. George and Andrey Kuzmichev and Gabriel Y. Lai and Jocelyn A. Lee and Ranell L. Myles and Shakira M. Nelson},
	Journal = {Preventive Medicine},
	Title = {Design and analysis of group-randomized trials in cancer- A review of current practices},
	Year = {2018},
	Volume = {111},
	Pages = {241-247}}

@article{turner17b,
	Author = {Elizabeth L. Turner and Melanie Prague and John A. Gallis and Fan Li and David M. Murray},
	Journal = {Am J Public Health},
	Title = {Review of Recent Methodological Developments in Group-Randomized Trials: Part 2—Analysis},
	Year = {2017},
	Volume = {107},
	Number = {7}}

@article{turner17a,
	Author = {Elizabeth L. Turner and Fan Li and John A. Gallis and Melanie Prague and David M. Murray},
	Journal = {Am J Public Health},
	Title = {Review of Recent Methodological Developments in Group-Randomized Trials: Part 1—Design},
	Year = {2017},
	Volume = {107},
	Number = {6}}

@article{crepsi,
	Author = {Catherine M. Crespi},
	Journal = {Annual Review of Public Health},
	Title = {Improved Designs for Cluster Randomized Trials},
	Year = {2016},
	Volume = {37},
	Number = {1},
	Pages = {1-16}}

@article{crt_comparative_bellamy,
	Author = {Scarlett L. Bellamy and Robert Gibberd and Lynne Hancock and Peter Howley and Bruce Kennedy and Neil Klar and Stuart Lipsitz and Louise Ryan},
	Journal = {Statistical Methods in Medical Research},
	Title = {Improved Designs for Cluster Randomized Trials},
	Year = {2000},
	Volume = {9},
	Pages = {135-159}}

@book{pearl,
	Author = {Judea Pearl},
	Edition = {2nd edition},
	Publisher = {Cambridge University Press},
	Title = {Causality: models, reasoning and inference},
	Year = {2009}}

@article{ku_protocol,
	Author = {Phelgona Otieno and Peter Waiswa and Elizabeth Butrick and Gertrude Namazzi and Kevin Achola and Nicole Santos and Ryan Keating and Felicia Lester and Dilys Walkter},
	Journal = {Annual Review of Public Health},
	Title = {Strengthening intrapartum and immediate newborn care to reduce morbidity and mortality of preterm infants born in health facilities in Migor County, Kenya and Busoga Region, Uganda: a study protocol for a randomized control trial},
	Year = {2018},
	Volume = {19},
	Number = {313}}

@article{rct_rosenblum_2010,
   author = {Michael Rosenblum and Mark van der Laan},
   title = {Simple, Efficient Estimators of Treatment Effects in Randomized Trials Using Generalized Linear Models to Leverage Baseline Variables},
   journal = {The International Journal of Biostatistics},
   volume = {6},
   number = {1},
   year = {2010},
   type = {Journal Article}
}

@article{superlearner_2007,
   author = {van der Laan, Mark and Polley, Eric C.  and Hubbard, Alan E. },
   title = {Super Learner},
   journal = {Statistical Applications in Genetics and Molecular Biology},
   volume = {6},
   number = {1},
   year = {2007},
   type = {Journal Article}
}

@article{rubin_vdl_max_eff_2008,
   author = {Daniel B Rubin and Mark J van der Laan},
   title = {Empirical efficiency maximization: improved locally efficient covariate adjustment in randomized experiments and survival analysis},
   journal = {The International Journal of Biostatistics},
   volume = {4},
   number = {1},
   year = {2008},
   type = {Journal Article}
}

@article{neyman1923,
   author = {Jerzy Neyman},
   title = {Sur les applications de la theorie des probabilites aux experiences agricoles: essai des principes (In Polish). English translation by D.M. Dabrowska and T.P. Speed (1990)},
   journal = {Statistical Science},
   volume = {5},
   pages = {463-472},
   year = {1923},
   type = {Journal Article}
}

@article{rubin1990,
   author = {Donald B. Rubin},
   title = {Comment: Neyman (1923) and causal inference in experiments and observational studies.},
   journal = {Statistical Science},
   volume = {5},
   number = {4},
   pages = {472-480},
   year = {1990},
   type = {Journal Article}
}

@article{imbens_2004,
   author = {Guido W. Imbens},
   title = {Nonparametric estimation of average treatment effects under exogeneity: a review},
   journal = {Review of Economics and Statistics},
   volume = {86},
   number = {1},
   pages = {4-29},
   year = {2004},
   type = {Journal Article}
}

@article{imai_2008,
   author = {Kosuke Imai},
   title = {Variance identification and efficiency analysis in randomized experiments under the matched-pair design},
   journal = {Statistics in Medicine},
   volume = {27},
   pages = {4857-4873},
   year = {2008},
   type = {Journal Article}
}

@article{balzer_sate,
   author = {Laura B. Balzer and Maya Petersen and Mark J. van der Laan and the SEARCH Collaboration},
   title = {Targeted estimation and inference for the sample average treatment effect in trials with and without pair-matching},
   journal = {Statistics in Medicine},
   volume = {35},
   pages = {3717-3732},
   year = {2016},
   type = {Journal Article}
}

@article{diaz_missing,
   author = {Karla Diaz-Ordaz and Michael G. Kenward and Abie Cohen and Claire L. Coleman and Sandra Eldridge},
   title = {Are missing data adequately handled in cluster randomised trials? A systematic review and guidelines},
   journal = {Clinical Trials},
   volume = {11},
   number = {5},
   pages = {590-600},
   year = {2014},
   type = {Journal Article}
}

@article{fiero,
   author = {Mallorie H. Fiero and Shuang Huang and Eyal Oren and Melanie L. Bell},
   title = {Statistical analysis and handling of missing data in cluster randomized trials: a systematic review},
   journal = {Trials},
   volume = {17},
   number = {72},
   year = {2016},
   type = {Journal Article}
}

@article{imai2009,
   author = {Kosuke Imai and Gary King and Clayton Nall},
   title = {The essential role of pair matching in cluster-randomized experiments, with application to the Mexican Universal Health Insurance Evaluation.},
   journal = {Statistical Science},
   volume = {24},
   number = {1},
   pages = {29-53},
   year = {2009},
   type = {Journal Article}
}

@article{balzer_matching,
   author = {Laura B. Balzer and Maya L Petersen and Mark J van der Laan and SEARCH Consortium},
   title = {Adaptive pair-matching in randomized trials with unbiased and efficient effect estimation},
   journal = {Statistics in Medicine},
   volume = {34},
   number = {6},
   pages = {999-1011},
   year = {2015},
   type = {Journal Article}
}

@article{pronto,
   author = {Dilys Walker and Susanna R. Cohen and Jimena Fritz and Marisela Olvera-García and Sarah T. Zelek and Jenifer O. Fahey and Martín Romero-Martínez and Alejandra Montoya-Rodríguez and Héctor Lamadrid-Figueroa},
   title = {Impact Evaluation of PRONTO Mexico: A Simulation-Based Program in Obstetric and Neonatal Emergencies and Team Training},
   journal = {Simulation in Healthcare},
   volume = {11},
   number = {1},
   pages = {1-9},
   year = {2015},
   type = {Journal Article}
}

@Manual{nbp_match,
    title = {nbpMatching: Functions for Optimal Non-Bipartite Matching},
    author = {Cole Beck and Bo Lu and Robert Greevy},
    year = {2016},
    note = {R package version 1.5.1},
    url = {https://CRAN.R-project.org/package=nbpMatching},
  }
